%% file: main.tex
\documentclass[twocolumn,epjc3]{svjour3}          

\RequirePackage[T1]{fontenc}
\smartqed  
\usepackage[utf8]{inputenc}
\RequirePackage{graphicx}
\RequirePackage{mathptmx}      
\RequirePackage{flushend}
\RequirePackage[numbers,sort&compress]{natbib}
\RequirePackage[colorlinks,citecolor=blue,urlcolor=blue,linkcolor=blue]{hyperref}
\usepackage[leftcaption]{sidecap}
\usepackage{graphics}
\pdfoutput=1
\input{basic/my_ds_macro.tex}
\usepackage{xcolor}
\usepackage{xurl}
\usepackage{float}
\usepackage{testhyphens}
\usepackage{hyphenat}
\input{basic/aria.def}
\input{basic/inst}
\usepackage{mathptmx}

\makeatletter
\renewcommand{\thanksref}[1]{\nolinebreak\textsuperscript{\ref{#1}}\nolinebreak\checknextarg}
\newcommand{\checknextarg}{\@ifnextchar\bgroup{\nolinebreak\gobblenextarg}{}}
\newcommand{\gobblenextarg}[1]{ \textsuperscript{\nolinebreak\hspace{-4pt}\mbox{\nolinebreak$^,$\nolinebreak\ref{#1}\nolinebreak}\nolinebreak} \@ifnextchar\bgroup{\gobblenextarg}{}}

\begin{document}


\title{Separating \ce{^39Ar} from \ce{^40Ar}   by cryogenic distillation with Aria for dark matter searches}
\author{The DarkSide-20k Collaboration$^\text{\normalfont a,1}$}
\thankstext{e1}{e-mail: ds-ed@lngs.infn.it}
\institute{See back for author list \label{addr1}}

%


\date{\today}

\maketitle
\begin{abstract}
    \input{sections/abstract}
\end{abstract}

\input{sections/intro}

\input{sections/requirements}

\input{sections/plant}

\input{sections/column}

\input{sections/shaft}

\input{sections/leak_test}

\input{sections/prototype}

\input{sections/performance}

\input{sections/Conclusion}

\input{sections/Acknoledgment}




\bibliographystyle{basic/epj}


\input{basic/author_list_end}

\end{document}

%% file: basic/my_ds_macro.tex
\usepackage[separate-uncertainty,retain-explicit-plus,per-mode=symbol,binary-units]{siunitx}
\usepackage[version=4]{mhchem}

\newcommand{\reffig}[1]{Fig.~\ref{fig:#1}}
\newcommand{\reffiginitpar}[1]{ Fig.~\ref{fig:#1}}
\newcommand{\reftab}[1]{Table~\ref{tab:#1}}
\newcommand{\refeqn}[1]{eq.~(\ref{eq:#1})}
\newcommand{\refsec}[1]{Sect.~\ref{sec:#1}}

\usepackage{enumitem}
\setdescription{itemsep=0pt,parsep=0pt,labelsep=.2cm,leftmargin=2.2cm,labelwidth=2cm,listparindent=.7cm}
\setlist{nolistsep,leftmargin=1cm}
\newlist{enumcompactitem}{itemize}{3}
\setlist[enumcompactitem]{topsep=0pt,partopsep=0pt,itemsep=0pt,parsep=0pt}
\setlist[enumcompactitem,1]{label=\textbullet}
\setlist[enumcompactitem,2]{label=---}
\setlist[enumcompactitem,3]{label=*}
\newlist{enumcompactdesc}{description}{3}
\setlist[enumcompactdesc]{topsep=0pt,partopsep=0pt,itemsep=0pt,parsep=0pt}
\newlist{enumcompactenum}{enumerate}{3}
\setlist[enumcompactenum]{topsep=0pt,partopsep=0pt,itemsep=0pt,parsep=0pt}
\setlist[enumcompactenum,1]{label=\arabic*}
\setlist[enumcompactenum,2]{label=\alph*}
\setlist[enumcompactenum,3]{label=\roman*}
\DeclareSIUnit\c{\mbox{$c$}}
\DeclareSIUnit\magn{\mbox{$\times$}}
\DeclareSIUnit\min{min}
\DeclareSIUnit\week{week}
\DeclareSIUnit\month{mo}
\DeclareSIUnit\months{mos}
\DeclareSIUnit\year{yr}
\DeclareSIUnit\years{years}
\DeclareSIUnit\yr{yr}
\DeclareSIUnit\standard{std}
\DeclareSIUnit\str{sr}
\DeclareSIUnit\ppm{ppm}
\DeclareSIUnit\ppb{ppb}
\DeclareSIUnit\ppt{ppt}
\DeclareSIUnit\pe{PE}
\DeclareSIUnit\spe{SPE}
\DeclareSIUnit\pdm{PDM}
\DeclareSIUnit\ev{events}
\DeclareSIUnit\ct{counts}
\DeclareSIUnit\neutron{\mbox{$n$}}
\DeclareSIUnit\smp{samples}
\DeclareSIUnit\Sample{S}
\DeclareSIUnit\ch{ch}
\DeclareSIUnit\hit{hit}
\DeclareSIUnit\hits{hits}
\DeclareSIUnit\bin{(\mbox{5-PE}~bin)}
\DeclareSIUnit\sgm{\mbox{$\sigma$}}
\DeclareSIUnit\rms{RMS}
\DeclareSIUnit\keVee{\mbox{keV$_{e{\rm e}}$}}
\DeclareSIUnit\keVr{\mbox{keV$_{\rm nr}$}}
\DeclareSIUnit\eVee{\mbox{eV$_{\rm ee}$}}
\DeclareSIUnit\eVr{\mbox{eV$_{\rm nr}$}}
\DeclareSIUnit\ph{photon}
\DeclareSIUnit\el{\mbox{$e^-$}}
\DeclareSIUnit\pm{\mbox{PMT}}
\DeclareSIUnit\pixel{\mbox{pixel}}
\DeclareSIUnit\inch{''}
\DeclareSIUnit\foot{'}
\DeclareSIUnit\bit{bit}
\DeclareSIUnit\sample{samples}
\DeclareSIUnit\barn{barn}
\DeclareSIUnit\bara{bar}
\DeclareSIUnit\barg{barg}
\DeclareSIUnit\mlardepth{\mbox(meter~of~\LAr~depth)}
\DeclareSIUnit\Curie{Ci}
\DeclareSIUnit\psf{psf}
\DeclareSIUnit\pcf{pcf}
\DeclareSIUnit\parsec{pc}
\DeclareSIUnit\mwe{\mbox{m.w.e.}}
\DeclareSIUnit\liveday{\mbox{live-days}}
\DeclareSIUnit\days{\mbox{days}}
\DeclareSIUnit\miles{\mbox{miles}}
\DeclareSIUnit\lumens{\mbox{lm}}
\DeclareSIUnit\degreeC{\mbox{$^{\circ}$C}}
\DeclareSIUnit\degreeF{\mbox{$^{\circ}$F}}
\DeclareSIUnit\electron{\mbox{$e^-$}}
\DeclareSIUnit\Euro{\mbox{\euro}}
\DeclareSIUnit\cph{cph}
\DeclareSIUnit\neq{neq}
\DeclareSIUnit\normal{\mbox{N}}

\newcommand{\AArArThreeNineOverArFourZeroRatio}{\num{8E-16}}
\newcommand{\AArArThreeNineActivity}{\SI{1}{\becquerel\per\kg}}

\newcommand{\DS}{\mbox{DarkSide}}

\newcommand{\DSf}{\mbox{DarkSide-50}}

\newcommand{\DSk}{\mbox{DarkSide-20k}}

\newcommand{\DSl}{\mbox{DarkSide-LowMass}}

\newcommand{\Argo}{\mbox{Argo}}

\newcommand{\GADMC}{\mbox{GADMC}}

\newcommand{\DEAP}{\mbox{DEAP-3600}}
\newcommand{\mCLEAN}{\mbox{MiniCLEAN}}
\newcommand{\ArDM}{\mbox{ArDM}}

\newcommand{\Urania}{\mbox{Urania}}
\newcommand{\Aria}{\mbox{Aria}}

\newcommand{\LArTPC}{\mbox{LAr~TPC}}

\newcommand{\LAr}{\ce{LAr}}

\newcommand{\AAr}{\ce{AAr}}
\newcommand{\UAr}{\ce{UAr}}

\newcommand{\LArNormalTemperature}{\SI{87}{\kelvin}}

\newcommand{\DSkTotalMass}{\SI{51.1}{\tonne}}

\newcommand{\DSlApproxMassScale}{\SI{1}{\tonne}}

\newcommand{\UraniaUArRate}{\SI{330}{\kg\per\day}}  

\newcommand{\AriaDepletionPerPass}{\num{10}}

\newcommand{\AriaSeruciHeight}{\SI{350}{\m}}

\newcommand{\AriaMonteSinniDiameter}{\SI{5}{\m}}

\newcommand{\AriaModulesNumber}{\num{30}}
\newcommand{\AriaCentralModulesNumber}{\num{28}}
\newcommand{\AriaCentralModulesHeight}{\SI{12}{\m}}

\newcommand{\AriaPlatformVerticalSpacing}{\SI{4}{\m}}

\newcommand{\ArgoTotalMass}{\SI{400}{\tonne}}

\newcommand{\DSfUArDepletionWithUncertainty}{$1400\pm 200$}

\newcommand{\DSfUArMassApprox}{\SI{150}{\kg}}

%% file: basic/inst.tex
\newcommand{\Alberta}{Department of Physics, University of Alberta, Edmonton, AB T6G 2R3, Canada}
\newcommand{\APC}{APC, Universit\'e de Paris, CNRS, Astroparticule et Cosmologie, Paris F-75013, France}
\newcommand{\AQLNGS}{INFN Laboratori Nazionali del Gran Sasso, Assergi (AQ) 67100, Italy}
\newcommand{\AQGSSI}{Gran Sasso Science Institute, L'Aquila 67100, Italy}

\newcommand{\AstroCeNT}{AstroCeNT, Nicolaus Copernicus Astronomical Center of the Polish Academy of Sciences, 00-614 Warsaw, Poland}
\newcommand{\Augustana}{Physics Department, Augustana University, Sioux Falls, SD 57197, USA}
\newcommand{\Belgorod}{Radiation Physics Laboratory, Belgorod National Research University, Belgorod 308007, Russia}

\newcommand{\BINP}{Budker Institute of Nuclear Physics, Novosibirsk 630090, Russia}
\newcommand{\BNLaddress}{Brookhaven National Laboratory, Upton, NY 11973, USA}
\newcommand{\BOINFN}{INFN Bologna, Bologna 40126, Italy}
\newcommand{\BOUniPHY}{Physics Department, Universit\`a degli Studi di Bologna, Bologna 40126, Italy}
\newcommand{\CS}{CarboSulcis S.p.A. - Miniera Monte Sinni, Cortoghiana 09010, Italy}
\newcommand{\CAUniCHE}{Department of Mechanical, Chemical, and Materials Engineering, Universit\`a degli Studi, Cagliari 09123, Italy}
\newcommand{\CAUniEEE}{Department of Electrical and Electronic Engineering, Universit\`a degli Studi di Cagliari, Cagliari 09123, Italy}
\newcommand{\CAUniPHY}{Physics Department, Universit\`a degli Studi di Cagliari, Cagliari 09042, Italy}
\newcommand{\CAINFN}{INFN Cagliari, Cagliari 09042, Italy}
\newcommand{\Carleton}{Department of Physics, Carleton University, Ottawa, ON K1S 5B6, Canada}

\newcommand{\CentroFermi}{Museo della fisica e Centro studi e Ricerche Enrico Fermi, Roma 00184, Italy}
\newcommand{\CERNaddress}{CERN, European Organization for Nuclear Research 1211 Geneve 23, Switzerland, CERN}
\newcommand{\CIEMAT}{CIEMAT, Centro de Investigaciones Energ\'eticas, Medioambientales y Tecnol\'ogicas, Madrid 28040, Spain}

\newcommand{\CPPM}{Centre de Physique des Particules de Marseille, Aix Marseille Univ, CNRS/IN2P3, CPPM, Marseille, France}
\newcommand{\CTINFN}{INFN Catania, Catania 95121, Italy}
\newcommand{\CTUNI}{Universit\`a of Catania, Catania 95124, Italy}
\newcommand{\CTLNS}{INFN Laboratori Nazionali del Sud, Catania 95123, Italy}
\newcommand{\ENEA}{Department of Fusion and Nuclear Safety Technologies, ENEA, Frascati 00044, Italy} 
\newcommand{\ENUniCEE}{Engineering and Architecture Faculty, Universit\`a di Enna Kore, Enna 94100, Italy}
\newcommand{\ETHZ}{Institute for Particle Physics, ETH Z\"urich, Z\"urich 8093, Switzerland}
\newcommand{\FNALaddress}{Fermi National Accelerator Laboratory, Batavia, IL 60510, USA}
\newcommand{\FortLewis}{Department of Physics and Engineering, Fort Lewis College, Durango, CO 81301, USA}
\newcommand{\GEUni}{Physics Department, Universit\`a degli Studi di Genova, Genova 16146, Italy}
\newcommand{\GEINFN}{INFN Genova, Genova 16146, Italy}
\newcommand{\Hawaii}{Department of Physics and Astronomy, University of Hawai'i, Honolulu, HI 96822, USA}
\newcommand{\Houston}{Department of Physics, University of Houston, Houston, TX 77204, USA}
\newcommand{\IHEPaddress}{Institute of High Energy Physics, Beijing 100049, China}
\newcommand{\INCDTIM}{National Institute for Research and Development of Isotope and Molecular Technologies, Cluj-Napoca 400293,
Romania}

\newcommand{\JINR}{Joint Institute for Nuclear Research, Dubna 141980, Russia}
\newcommand{\Krakow}{M.~Smoluchowski Institute of Physics, Jagiellonian University, 30-348 Krakow, Poland}
\newcommand{\Kurchatov}{National Research Centre Kurchatov Institute, Moscow 123182, Russia}
\newcommand{\Laurentian}{Department of Physics and Astronomy, Laurentian University, Sudbury, ON P3E 2C6, Canada}
\newcommand{\Lancaster}{Physics Department, Lancaster University, Lancaster LA1 4YB, UK}
\newcommand{\LNFINFN}{INFN Laboratori Nazionali di Frascati, Frascati 00044, Italy}
\newcommand{\LNLINFN}{INFN Laboratori Nazionali di Legnaro, Legnaro (Padova) 35020, Italy}
\newcommand{\Lodz}{Institute of Applied Radiation Chemistry, Lodz University of Technology, 93-590 Lodz, Poland}
\newcommand{\LPNHE}{LPNHE, CNRS/IN2P3, Sorbonne Universit\'e, Universit\'e Paris Diderot, Paris 75252, France}

\newcommand{\Manchester}{Department of Physics and Astronomy, The University of Manchester, Manchester M13 9PL, UK}
\newcommand{\MEPhI}{National Research Nuclear University MEPhI, Moscow 115409, Russia}
\newcommand{\MendeleevUniverisity}{Mendeleev University of Chemical Technology, Moscow 125047, Russia}

\newcommand{\MIINFN}{INFN Milano, Milano 20133, Italy}
\newcommand{\MIPoliICA}{Civil and Environmental Engineering Department, Politecnico di Milano, Milano 20133, Italy}
\newcommand{\MIPoliCHE}{Chemistry, Materials and Chemical Engineering Department ``G.~Natta", Politecnico di Milano, Milano 20133, Italy}

\newcommand{\MIUni}{Physics Department, Universit\`a degli Studi di Milano, Milano 20133, Italy}
\newcommand{\MSU}{Skobeltsyn Institute of Nuclear Physics, Lomonosov Moscow State University, Moscow 119234, Russia}
\newcommand{\NAINFN}{INFN Napoli, Napoli 80126, Italy}
\newcommand{\NAUniPHY}{Physics Department, Universit\`a degli Studi ``Federico II'' di Napoli, Napoli 80126, Italy}
\newcommand{\NAUniCHE}{Chemical, Materials, and Industrial Production Engineering Department, Universit\`a degli Studi ``Federico II'' di Napoli, Napoli 80126, Italy}
\newcommand{\NAUniPHARM}{Pharmacy Department, Universit\`a degli Studi ``Federico II'' di Napoli, Napoli 80131, Italy}
\newcommand{\NAUniStruct}{Dipartimento di Strutture per l'Ingegneria e l'Architettura, Universit\`a degli Studi ``Federico II'' di Napoli, Napoli 80131, Italy}

\newcommand{\NSU}{Novosibirsk State University, Novosibirsk 630090, Russia}
\newcommand{\OACINAF}{INAF Osservatorio Astronomico di Capodimonte, 80131 Napoli, Italy}
\newcommand{\Petersburg}{Saint Petersburg Nuclear Physics Institute, Gatchina 188350, Russia}

\newcommand{\PIINFN}{INFN Pisa, Pisa 56127, Italy}
\newcommand{\PIUniPHY}{Physics Department, Universit\`a degli Studi di Pisa, Pisa 56127, Italy}
\newcommand{\PNNLaddress}{Pacific Northwest National Laboratory, Richland, WA 99352, USA}
\newcommand{\Princeton}{Physics Department, Princeton University, Princeton, NJ 08544, USA}
\newcommand{\Polaris}{Polaris S.r.l., Misinto 20826, Italy}
\newcommand{\Queens}{Department of Physics, Engineering Physics and Astronomy, Queen's University, Kingston, ON K7L 3N6, Canada}
\newcommand{\RHUL}{Department of Physics, Royal Holloway University of London, Egham TW20 0EX, UK}
\newcommand{\RMTreINFN}{INFN Roma Tre, Roma 00146, Italy}
\newcommand{\RMTreUni}{Mathematics and Physics Department, Universit\`a degli Studi Roma Tre, Roma 00146, Italy}
\newcommand{\RMUnoINFN}{INFN Sezione di Roma, Roma 00185, Italy}
\newcommand{\RMUnoUni}{Physics Department, Sapienza Universit\`a di Roma, Roma 00185, Italy}
\newcommand{\SAINFN}{INFN Salerno, Salerno 84084, Italy}
\newcommand{\SAUni}{Physics Department, Universit\`a degli Studi di Salerno, Salerno 84084, Italy}
\newcommand{\SNOLABaddress}{SNOLAB, Lively, ON P3Y 1N2, Canada}

\newcommand{\Temple}{Physics Department, Temple University, Philadelphia, PA 19122, USA}
\newcommand{\TNFBK}{Fondazione Bruno Kessler, Povo 38123, Italy}
\newcommand{\TNTIFPA}{Trento Institute for Fundamental Physics and Applications, Povo 38123, Italy}

\newcommand{\TOINFN}{INFN Torino, Torino 10125, Italy}
\newcommand{\TOPoli}{Department of Electronics and Communications, Politecnico di Torino, Torino 10129, Italy}

\newcommand{\TRIUMFaddress}{TRIUMF, 4004 Wesbrook Mall, Vancouver, BC V6T 2A3, Canada}

\newcommand{\UB}{Universiatat de Barcelona, Barcelona E-08028, Catalonia, Spain} 
\newcommand{\UCDavis}{Department of Physics, University of California, Davis, CA 95616, USA}

\newcommand{\UCLA}{Physics and Astronomy Department, University of California, Los Angeles, CA 90095, USA}
\newcommand{\UMass}{Amherst Center for Fundamental Interactions and Physics Department, University of Massachusetts, Amherst, MA 01003, USA}

\newcommand{\USP}{Instituto de F\'isica, Universidade de S\~ao Paulo, S\~ao Paulo 05508-090, Brazil}
\newcommand{\VTech}{Virginia Tech, Blacksburg, VA 24061, USA}
\newcommand{\WilliamsCollege}{Williams College, Physics Department, Williamstown, MA 01267 USA}
\newcommand{\Zaragoza}{Centro de Astropart\'iculas y F\'isica de Altas Energ\'ias, Universidad de Zaragoza, Zaragoza 50009, Spain}
\newcommand{\ZaragozaARAID}{Fundaci\'on ARAID, Universidad de Zaragoza, Zaragoza 50009, Spain}
\newcommand{\MISE}{Now at Ministero dello Sviluppo Economico, Palazzo Piacentini,
00187 Roma, Italy}

%% file: sections/abstract.tex
 The Aria project consists of a plant, hosting a \AriaSeruciHeight\ cryogenic isotopic distillation column, the tallest ever built, which is currently in the installation  phase in a mine shaft at  Carbosulcis S.p.A., Nuraxi-Figus (SU), Italy. 
 \Aria\ is one of the pillars of the argon dark-matter search experimental program, lead by the Global Argon Dark Matter Collaboration. 
 \Aria\ was designed to reduce the isotopic abundance of \ce{^39Ar}, a $\beta$-emitter of cosmogenic origin, whose activity poses background and pile-up concerns in the detectors, in the argon used for the dark-matter searches, the so-called Underground Argon (\UAr). 
 In this paper, we discuss the requirements, design, construction,  tests, and projected performance   of the  plant for the isotopic cryogenic distillation of argon.
 We also present the successful results of  isotopic cryogenic distillation  of nitrogen  with a prototype plant,  operating the column at total reflux.

%% file: sections/intro.tex
\section{Introduction}
\label{sect:context}
Large liquid argon detectors offer one of the best avenues for detecting galactic Weakly Interacting Massive Particles (WIMPs) via their scattering on atomic nuclei.
However, atmospheric argon (AAr) has a  naturally occurring radioactive isotope, \ce{^39Ar}, of isotopic abundance of \AArArThreeNineOverArFourZeroRatio\ in mass, which is a $\beta$-emitter of cosmogenic origin, and whose activity of about  \AArArThreeNineActivity\ raises background and pile-up concerns.
Indeed, the liquid argon target allows for  powerful discrimination between nuclear and electron recoil scintillation signals via pulse-shape discrimination~\cite{Agnes:2015gu,Agnes:2016fz,Amaudruz:2018gr}, provided the background rate is not too high. However, this discrimination method cannot be applied in   experiments that look at the ionization signal only~\cite{Agnes:2018fg,Agnes:2018ft}.
The use of argon extracted from underground wells, the so-called Underground Argon  (\UAr), with a highly suppressed  content  of \ce{^39Ar}, is therefore pivotal for the physics potential of  dark matter search experiments. 

The \DSf\ experiment,  a liquid argon time projection chamber (\LArTPC) at Laboratori Nazionali del Gran Sasso (LNGS),  used a \DSfUArMassApprox\ active mass of \UAr\, extracted from \ce{CO2} wells  in Cortez, CO, USA, and measured the \ce{^39Ar} Depletion Factor ($DF$)  with respect to AAr  to be \DSfUArDepletionWithUncertainty~\cite{Agnes:2016fz}. 
 A new production chain, that was recently set up to significantly increase the production of \UAr. This new production needs to meet the target requirements of the Global Argon Dark Matter Collaboration (\GADMC), a  worldwide effort that unifies the  \DS, \DEAP, \mCLEAN, and \ArDM\ experimental groups, for the construction of new experiments for argon dark-matter searches. In order of increasing size, these new  experiments are a potential  Dark\-SideLowMass, with approximately \DSlApproxMassScale\ target,  optimized for the detection of low-mass dark matter, aiming at improving the  world-leading results  of the \DSf\ experiment~\cite{Agnes:2018fg,Agnes:2018ft}, the  \DSkTotalMass\ target mass
  \DSk~\cite{Aalseth:2018gq},  under construction at LNGS, Italy, and the prospected  \Argo, of  \ArgoTotalMass\ target mass, that  will push the experimental sensitivity down to the so-called  neutrino floor.
The  argon procurement for this new production chain starts from the \Urania\ plant, now in the construction phase in Cortez, CO, USA,  that will extract and purify    \UAr\ at a maximum production rate of about \UraniaUArRate.
The \ce{^39Ar} radioactivity of \UAr, though remarkably lower than that of \AAr, is neither low enough for the needs of the \DSl\ experiment, where it would be the limiting background to the cross-section sensitivity, nor for  the \Argo\ experiment, where  it would  cause   a major event pile-up, if built with the double-phase TPC technology.

The cryogenic isotopic distillation plant \Aria, which is currently in the installation  phase in a mine shaft  at CarboSulcis S.p.A., in Nuraxi-Figus (SU), Italy,  was designed to further reduce  the \ce{^39Ar} isotopic fraction of the \UAr\ by another factor of \AriaDepletionPerPass\ per pass, with a production rate of several \SI{}{\kg\per\day}. 
While the \AriaSeruciHeight\ tall, \AriaColumnInnerDiameter\ inner diameter, distillation column under construction fits  the needs of DarkSide-Low\-Mass in terms of production rate, for the larger \Argo\ experiment  a new wider column would need to be built. 

Cryogenic isotopic distillation with rectifying columns is a well established technique \cite{ANDREEV2007247} and has received quite some attention in the context of the stable isotope separation of the main biogenic elements such as carbon and oxygen and some industrial-scale  plants have already been built.
However, for argon isotopic distillation, this is the first time that such a plant is proposed and constructed.
In addition to cryogenic distillation, a few other techniques are currently  available for the separation of argon isotopes, based on the difference in molecular mass, such as centrifugal separation    and  diffusion separation, the latter based on the different average speed, at  thermal equilibrium, among isotopes of the same energy.
However, their application is limited by the low  yield and the  high cost per unit mass of separated isotopes.
As a matter of fact, the cryogenic isotopic distillation plant \Aria\ appears as a very promising new avenue   for the  depletion from \ce{^39Ar}  of such   large quantities of  argon,  at  reasonable cost and time. 
It is interesting to note    that target purification via distillation, though not isotopic, with cryogenic columns in the context of  dark matter search detectors was also pursued by other  collaborations using xenon \cite{Wang2014,2014RScI...85a5116W,ABE2009290,xenon}.  

The technological capability to achieve efficient isotopic separation with cryogenic distillation allows to widen the  application of the \Aria\ project to other fields, where the production of stable isotopes is  required, such as e.g. in medical applications.
However, in this paper we will focus on the application of the \Aria\ plant to the isotopic distillation of  argon.

A very important achievement for this project was  a nitrogen distillation run of the prototype plant,  a short version of the \Aria\ column using only the reboiler, the condenser and one central module, together with all the auxiliary equipment of the full column, installed in a surface building. The successful outcome of this run paved the way to the continuation of the project and the construction of the full plant.

%% file: sections/requirements.tex
\section{Design requirements}
\label{sec:requirements}
Isotopic separation by cryogenic distillation exploits the relative  volatility  of different isotopes, namely, for ideal mixtures,  the ratio of their vapor pressures at a given temperature.
 Continuous distillation, with a large number of  distillation stages, where the liquid and vapor phases undergo a countercurrent  exchange at thermodynamic equilibrium, is used to optimize the separation of isotopes that have  relative volatility close to unity.  
 As detailed in \refsec{plant}, heat is constantly provided from a bottom heat exchanger, in the so-called {\em reboiler}, that vaporizes the liquid, and extracted from a top heat exchanger, in the so-called {\em condenser}, that condenses the vapor.  To perform the isotopic separation,  the column temperature ranges between the boiling point of \ce{^40Ar} (bottom)  and of \ce{^39Ar} (top)  at the operating pressure, slightly above \SI{1}{\bar}. 
The pre-cooled \UAr\  feed  enters the column  at a given height and   flow.
The vapor rises in the column and re-condenses, while the liquid sinks by gravity and then reboils. 
In the rectifying section (above the feed point), the  molar  fraction of  \ce{^39Ar}  is larger than  in the feed argon, while in the stripping section (below the feed point) it is smaller  than in the feed argon.
Liquid argon depleted of  \ce{^39Ar}  is then  collected continuously from the bottom of the column, whereas  argon enriched of   \ce{^39Ar} is collected from the top.
Since the \ce{^39Ar} has a very  low isotopic fraction even in atmospheric  argon,  its  volatility relative to the other argon isotopes was never  measured. Therefore, for the column design, the relative volatility  of  \ce{^39Ar} to \ce{^40Ar}, or its  more commonly used natural logarithm, $\mathrm{ln(\alpha_{39-40})}$, was derived from the measured relative volatility  of  \ce{^36Ar} to \ce{^40Ar}, which is  
0.0060\,$\pm$\,0.0001~\cite{FIESCHI1961453},  at the mean operating temperature of the column of  \AriaMeanOperatingTemperature, as shown in \refsec{plant}.
According to the model of~\cite{CanongiaLopes:2003ju}, the dependence of $\mathrm{ln(\alpha_{A-40})}$ on the isotopic mass $A$ is  $\mathrm{ln(\alpha_{A-40})\propto (40-A)/A}$. This means that, at \AriaMeanOperatingTemperature, $\mathrm{ln(\alpha_{39-40})}$ is about \AriaArLogVolatiityRatioMean\ and $\alpha_{39-40}$ is about  \AriaArVolatiityRatioMean, which is the value assumed for all the calculations in this paper. The uncertainty coming from model extrapolation is however difficult to estimate and therefore this number should be taken as an approximate value. The temperature dependence of $\mathrm{\alpha_{39-40}}$ turns out to be  about -0.00005/K.

The relative volatility of two species provides an estimate of how difficult it is to separate two species. 
Since this ratio is close to 1 for \ce{^39Ar} and \ce{^40Ar}, the separation is expected to be very difficult.
To optimize the distillation process,  the  \Aria\ column  makes use of a  high performance  packing material instead of the  distillation trays. The two
related quantities that characterize the separation capability of  distillation  columns are  the number of equivalent theoretical trays, $N_T$, (in a distillation columns the $N$ theoretical stages consist of the $N_T$ theoretical trays with the eventual addition of the reboiler and the condenser, in the case they represent equilibrium stages) and the Height Equivalent to a Theoretical Plate, HETP, the product of the two yielding the total column height.

In the \Aria\ distillation column, for a binary mixture,  the minimum number of theoretical stages need\-ed, $N_{min}$, at total reflux  is given by    the Fenske equation \cite{Fenske:1932do}:

\begin{equation}
 N_{\mathrm{min}}=\frac{\mathrm{ln} (S^0_{39-40})}{\mathrm{ln}(\alpha_{39-40})},
 \label{eq:Fenske}
\end{equation}
 for a   separation, $S^0_{39-40}$: 
 \begin{equation}
 S^0_{39-40}=\frac{x_D}{1-x_D}\frac{1-x_B}{x_B}\sim \frac{x_D}{x_B}, 
 \label{eq:Fenske2}
\end{equation}
where $x_D$ is the molar fraction of   \ce{^39Ar} in the top, $x_B$ the  molar fraction of \ce{^39Ar} in the bottom, and 
 $x_D,x_B \ll 1$. 
 


 Requiring for instance a separation   of \AriaDepletionPerPass,  from \refeqn{Fenske} it follows that $N_{min}$=\AriaNInfiniteForNominalHeavyRecovery. 
Moreover, when the column operates in finite reflux mode,  the  number of required stages is  larger than $N_{min}$.
To include such a large number of stages, the column needs to be very tall 
and  be filled with  high performance packing, i.e. with a small HETP. Moreover, for efficient use of the packing,  there is a limitation on the liquid flow per unit area, usually specified by the vendor. Therefore not only the height but also the  diameter matters for sizeable distillate production.

To support such a tall column,  a convenient  and cheap solution was found  with its installation in  an underground vertical mineshaft of \AriaMonteSinniDiameter\  diameter and \AriaSeruciHeight\ depth, dug in the 1940s, which was made available to \Aria\ by  
 the end of the  mine coal extraction cycle,  at the end of  2018.
 
The first phase of the \Aria\ project, which is the subject of this paper, consists  of a column of internal diameter $d$=\AriaColumnInnerDiameter, with \AriaColumnThickness\ wall thickness, enclosed in a vacuum cold box of \AriaCentralModulesDiameter\ diameter, with a total height approximately equal to the mineshaft depth. 
The  support structure of the column in the shaft is designed in a way to allow for the installation at later times of a wider column with  a maximum cold box diameter of  \AriaSeruciTwoExternalDiameter.

The rest of the paper is organized as follows.
In \refsec{plant} we discuss the plant design, followed by a description of the column in \refsec{column}. In \refsec{leak} we present the column vacuum leak tests. In \refsec{prototype} we discuss the prototype tests and the validation of some characteristics of the plant with measurements,  and in \refsec{performance} the  projected performance of Aria for argon isotopic distillation.

%% file: sections/plant.tex
\section{Plant design}
\label{sec:plant}

\begin{figure*}
\centering
\includegraphics[width=0.8\textwidth]{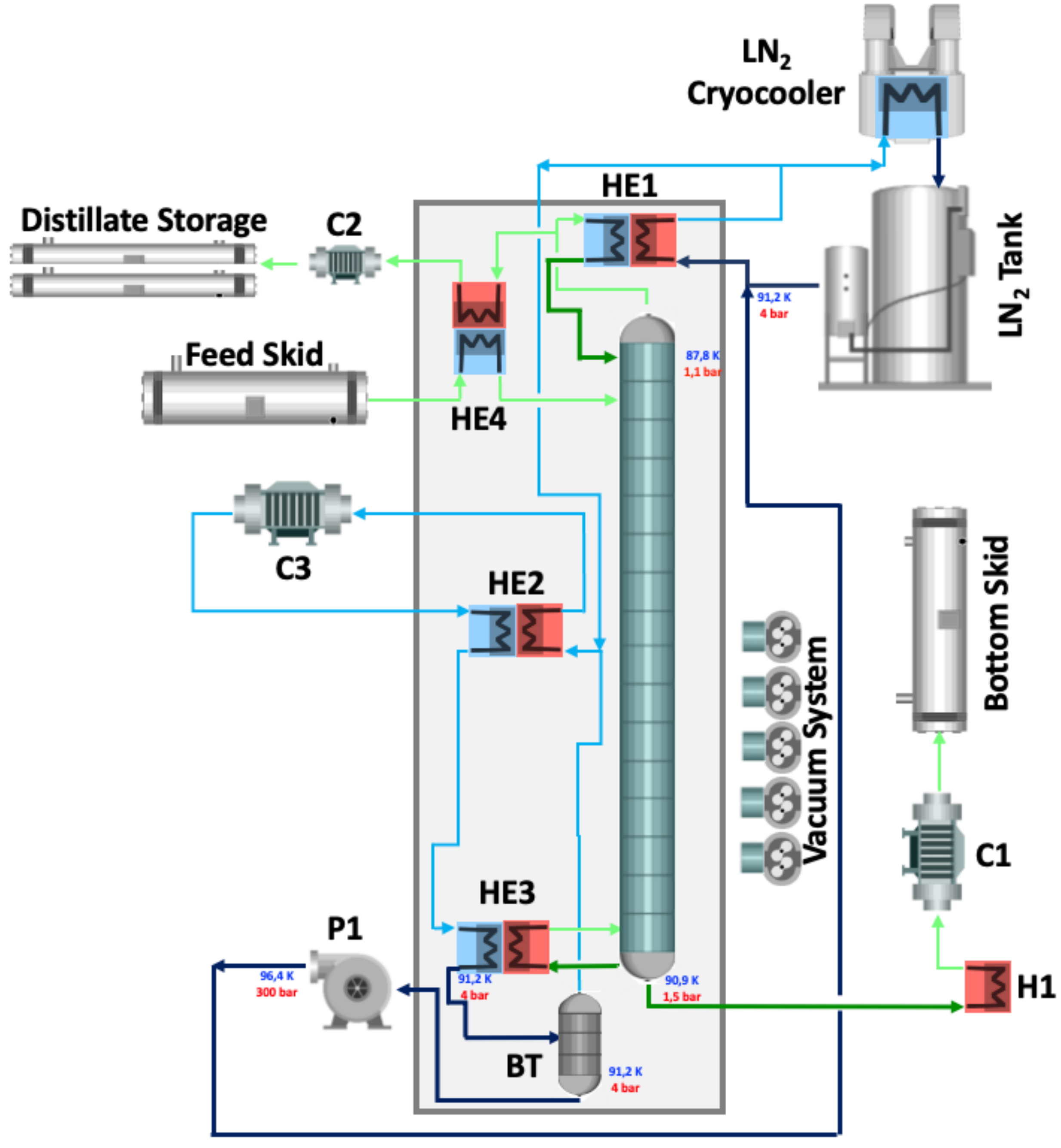}
\caption[]{Simplified scheme of the  \Aria\ plant. The full description can be found in the text. The color-coding of the heat exchangers is such that the red section provides heat to the fluid while the blue section removes heat from it. The scheme also reports  the values of operating pressure and temperature  for \ce{^39Ar}-\ce{^40Ar} distillation, as  obtained from a plant engineering  simulation (Aspen - HYSYS). 
}
\label{fig:Aria-Block-Diagram}
\end{figure*}

The \Aria\  plant  simplified scheme   is displayed in \reffig{Aria-Block-Diagram}. 
The column,  cryogenic tanks, and heat exchangers are enclosed in a cold box (grayed area) which is  vacuum-tight and designed to reduce thermal losses.
The cryogenic circuit of the plant  is designed  with  two independent loops: the  argon loop (dark green lines for the liquid and light green lines for the vapor/gas)
and the refrigeration loop, with  nitrogen gas (cyan lines) and liquid (dark blue lines) that are used to evaporate and to condense the argon. The \Aria\ plant was designed in a way that  minimizes nitrogen  consumption  and optimizes   energy efficiency, by using a closed-loop refrigeration circuit and appropriate use of
heat exchangers.

\UAr\  will be transported from the Urania plant being constructed in Cortez, Colorado, USA,  to  \Aria\ in Sardinia, Italy, and then from \Aria\ to LNGS, Italy,  inside  gas skids.
The argon gas from the Feed Skid is fed into the distillation column  through a  flow controller, and   pressure-regulated  to about \AriaColumnSimulationPressure. 
A  heat exchanger (HE4) with the  output distillate stream is used to cool the  argon.
The  bottom stream comes out of the  column as a liquid, gets heated as it passes through an air  heater (H1), compressed (C1) 
and  then delivered to the Bottom Skid. This is the argon that will be used in the  experiments.
At the top of the column, the distillate stream,  enriched in \ce{^39Ar}, is  delivered to the Distillate Storage after passing through HE4 and a compressor  (C2).

Liquid nitrogen  is used as  cooling fluid  in the  heat exchanger (HE1)  of the column condenser. The nitrogen vapor from the output of HE1  is heated through the  heat  exchanger HE2 and then  compressed, by a  screw rotary compressor (C3),  
to a pressure value between \SI{2}{\bar} and \SI{4}{\bar}. 
The compressed gas, after cooling in  HE2,  is used as heating fluid in the  heat exchanger (HE3) of the reboiler. The liquefied nitrogen, after passing through a nitrogen phase separator tank (BT),  is pumped by a  modular reciprocating pump  (P1), 
with a delivery pressure up to 
\AriaPumpedLiquidNitrogenPressure, all the way up to the top of the  column, and fed back to HE1. 
Liquid nitrogen is stored and fed into the circuit from  an external \AriaTankStorageVolume\ tank. The  excess  nitrogen gas from the system is  fed back to the tank, 
after being liquefied by four \AriaSingleStrilingPower\ cryogenerators (Stirling Cryogenics), inherited  from  the ICARUS experiment at LNGS.

Brazed plate heat exchangers 
are used for the reboiler (HE3), the condenser (HE1), and  HE2. These heat exchangers are characterized by high heat transfer efficiency and limited size  and are the ideal solution for this application. Coil heat exchangers (H1 and HE4) are used for the inlet and outlet argon flows.

\reffiginitpar{Aria-Block-Diagram} reports also the values of operating pressure and temperature, for \ce{^39Ar}-\ce{^40Ar} distillation, obtained with a plant engineering  simulation  using the Aspen HYSYS package. 
It can be seen that the column operating temperature varies from the top to the bottom  between \AriaMinOperatingTemperature\ and \AriaMaxOperatingTemperature.


%% file: sections/column.tex
\section{Column and cold box structure}
\label{sec:column}

\begin{figure}
\centering
\includegraphics[width=0.45\textwidth]{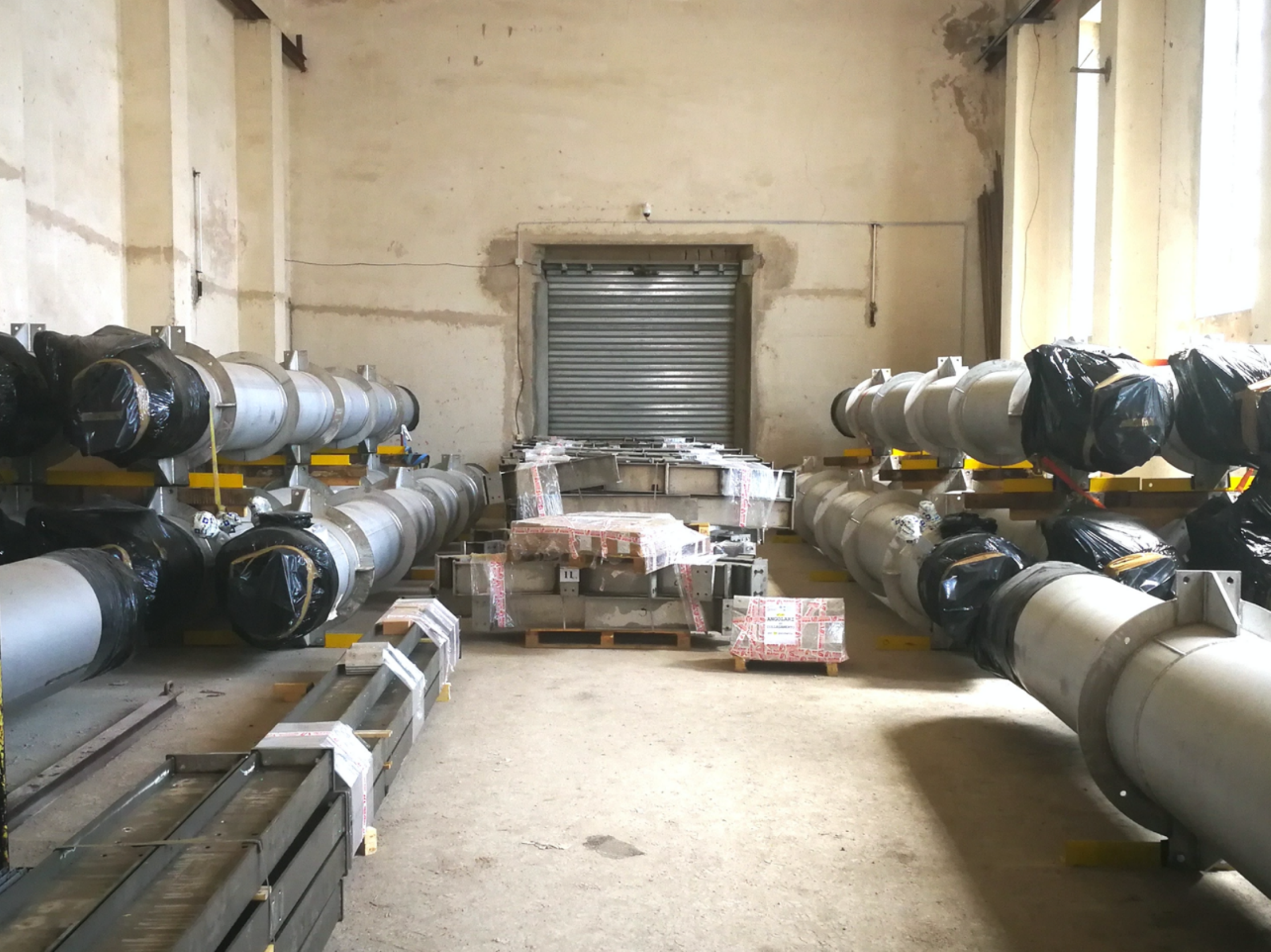}
\caption{The  central modules of the column stored at  Carbosulcis S.p.A., Nuraxi-Figus site, ready for installation.}
\label{fig:moduli_stored}
\end{figure}

\begin{figure}
\centering
\includegraphics[width=0.45\textwidth]{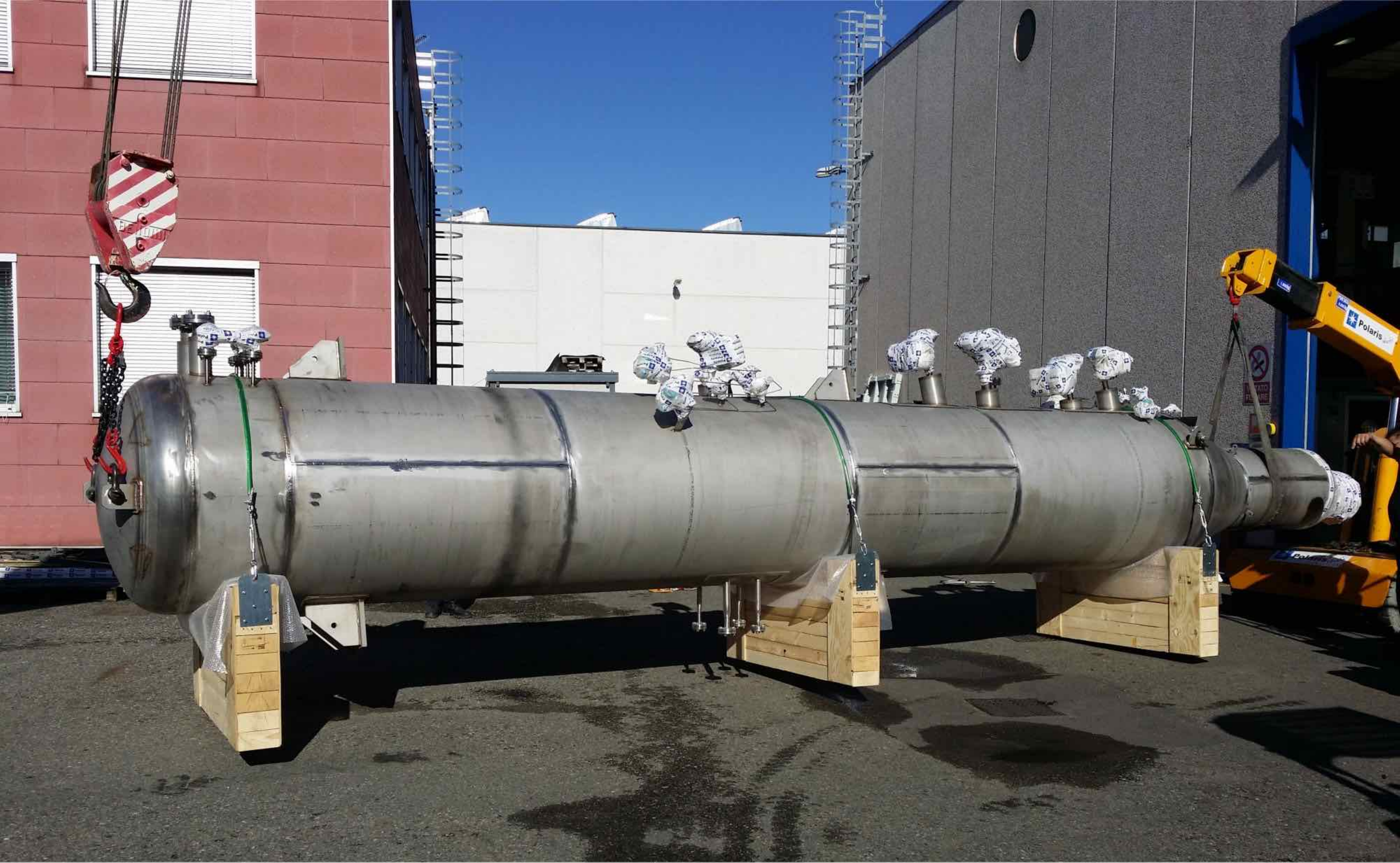}
\caption{The  top module of the column.}
\label{fig:top}
\end{figure}

\begin{figure}
\centering
\includegraphics[width=0.45\textwidth]{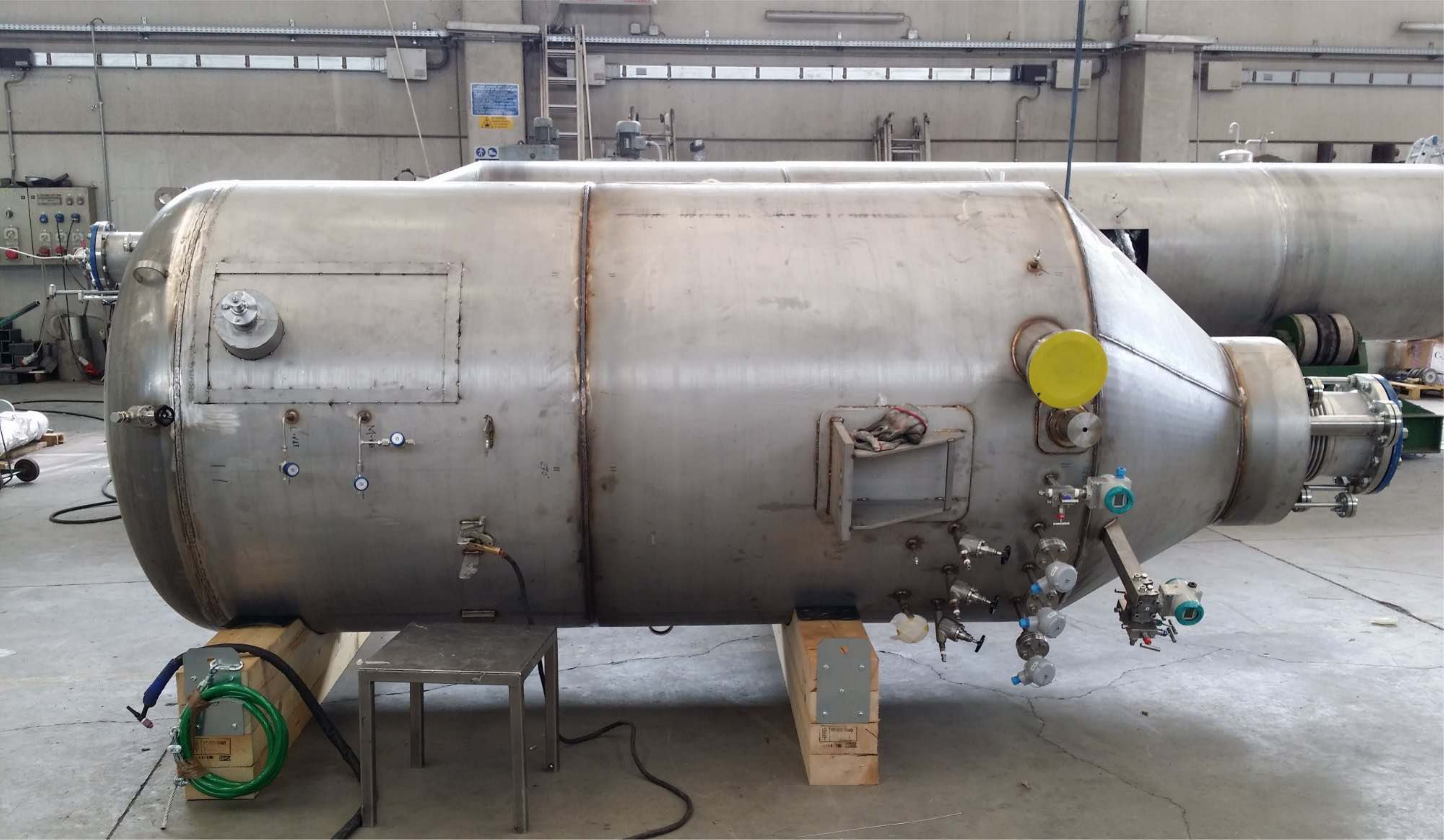}
\caption{The  bottom module of the column.}
\label{fig:bottom}
\end{figure}

For construction and transport, both the column and the surrounding cold box have  a modular structure. The thirty modules    were assembled  at the production site. 
The  
\AriaCentralModulesNumber\ central modules are identical cylindrical elements about \AriaCentralModulesHeight\ tall, with a \AriaCentralModulesDiameter\ diameter and an approximate weight of \AriaCentralModulesWeight.  
The top module, about \AriaTopModuleHeight\ tall and   \AriaTopModuleDiameter\ diameter,   hosts the top of the distillation column, about \AriaTopDistillationColumnHeight\ high, the condenser (HE1), a  liquid nitrogen buffer tank, not shown in the simplified scheme of \reffig{Aria-Block-Diagram}, 
and two heat recovery exchangers (HE4 and HE2). The bottom module, about 
\AriaBottomModuleHeight\ 
tall and 
\AriaBottomModuleDiameter\ 
diameter,  hosts the bottom of the distillation column, about \AriaTopDistillationColumnHeight\ high,  the reboiler (HE3), and a nitrogen phase separator tank (BT). 
\reffiginitpar{moduli_stored} shows some of the  central modules stored at the Carbosulcis site, ready for installation in the shaft. 
\reffiginitpar{top} displays the  top module while \reffig{bottom}  shows the   bottom module.

The  structure of the cold box, the internal equipment, and the piping are fully welded to reduce the risk of leaks. All weldings were  performed at the manufacturing company where  the modules were assembled, except for the orbital welds between modules, which will be performed in the mineshaft.

To account for the thermal contraction of the structure, axial bellows are interleaved between  every other module. At cold, the bellows expands  by about \SI{3}{\cm}. Due to the presence of bellows, the support of each module is independent from the others. The load is distributed laterally to the shaft walls. Every module is supplied with anchor points, whose sizing takes into consideration both the static weight and the stresses due to the cold box operating pressure. The anchor points are bolted to a structure, discussed in \refsec{shaft}, which is fixed to the lateral wall of the shaft. 

\subsection{Internal structure}

\begin{figure}
\centering
\includegraphics[width=0.45\textwidth]{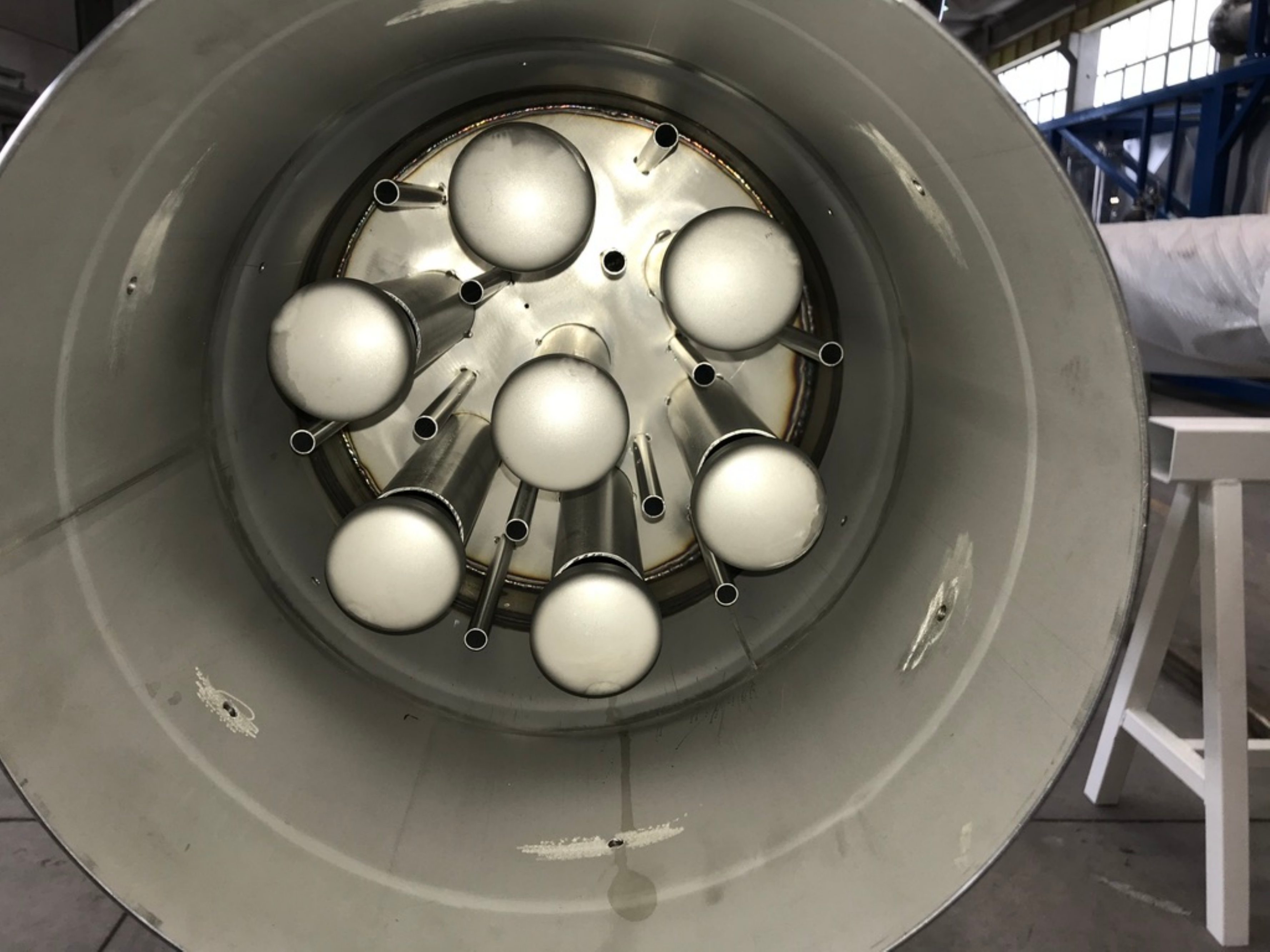}
\caption{A view from above of a liquid distributor.}
\label{fig:liquid}
\end{figure}

The \AriaCentralModulesNumber\  central modules  are filled with a structured  stainless steel packing  (Sulzer CY gauze).
To stay below the flooding limit and, therefore, guarantee an efficient distillation  with this packing,  according to measurements performed by the vendor with chloro\-benzene/\-etil\-benzene mixtures, a maximum  specific liquid volume flow rate or load, $\hat{V}_{L}$,  of \AriaLiquidFlowForTheoreticalPlate, is allowed.  Given the column inner diameter of \AriaColumnInnerDiameter, this  corresponds to a  liquid volume flow rate, $V_{L}$, of \AriaMaxLiquidFlow\ or to a mass flow rate of \AriaMaxMassLiquidFlow. 
With this load  and at the  average operating  pressure of the column with argon, the sizing parameter for packed columns $F$,  defined as $U_V\cdot \sqrt{\rho_V}$, where $U_V$ is the superficial gas velocity, i.e. the vapor volume flow rate, $V_{V}$, per unit  column cross section, and $\rho_V$ the  vapor argon density at equilibrium, whose value for argon is given in \reftab{input_parameters}, is \AriaFValueArgonNominal.
With this value of $F$,  from the reference  curves provided by the packing  producer (Sulzer) and approximating to  atmospheric pressure, one would expect an  HETP of about  \AriaHeightForTheoreticalPlate\ and a pressure drop of \AriaPressurDropForTheoreticalPlateNew. 
The measurement  of these parameters in the cryogenic environment is  an essential step of this research, and  is the main focus of the tests described in \refsec{prototype}. 
To avoid the channeling of the fluid in the packing and to optimize the uniformity across the column section,  each module  is divided into four sub-sections of packing, with an active height of \AriaColumnModuleSectionHeight\ each, interleaved with  a liquid distributor, shown in \reffig{liquid}.
The liquid formed on the distributor plate is streamed, through holes located at \AriaColumnDistributorLiquidHeight\ height in  perforated pipes uniformly distributed along the plate surface,   to the packing section below.  The vapor rises towards the packing section above through \AriaColumnDistributorVaporHeight\ high chimneys.
The total active height of the column is about \AriaColumnActiveHeight, that  corresponds to a   $N$ of \AriaNumberOfTheorecticalStages.
The pressure drop along the column is about 0.7 bar, with 0.5 bar due to  the   distributors.
The 
minimum argon mass that needs to be in the column for efficient distillation, is largely dominated by the liquid component, the vapor contributing only to  \AriaVaporContributionToHoldUp\ of the total. The two major contributions 
come    from the distributors, \AriaLiquidDistributorContributionToHoldUp, and from the wetting  of the packing, the so called holdup,  \AriaLiquidPackingContributionToHoldUp. The packing wettability was assumed for this calculation to be  \AriaPackingWettingFraction, as specified by the packing vendor. Again, its value  was  given for the above mentioned organic mixture  and, therefore, will  need to be verified at the cryogenic temperatures of the column with argon.  The total argon mass in the column during distillation, with the above mentioned assumptions, turns out to be approximately \AriaHoldUp. 

The thermal load of the column was calculated 
assuming the
maximum  liquid flow specified by  the packing producer, as   discussed in \refsec{column}.
The required thermal duty for the cryogenic system  turns out  about \AriaThermalDuty, broadly given  by the maximum liquid flow times the heat of vaporisation. 
The total  electric power needed	for the plant operation is about \AriaElectricPower, including the cryocooler, compressors, fluid, and vacuum system pumps load.





\subsection{Thermal insulation}

To minimize heat transfer through  conduction and radiation from the environment   to the cryogenic distillation column, a \AriaColdBoxVacuum\ vacuum is made in the cold box. In order to maintain the desired vacuum level, several pump stations of total pumping speed   \AriaVacuumPumpingSpeed, are installed along the column. In addition,   \AriaNumberofMLIColumn\  layers of Multi-Layer Insulation (MLI) are wrapped around  the column,  and \AriaNumberofMLILines\ are wrapped around all the other lines and reservoirs within the col box. With this  insulation, the residual thermal radiation input power to the column is about \AriaRadiationInputPowerPerSquareMeter, a few percent  of the thermal duty cycle of the column. 
Insulation is also  provided  on the equipment and piping outside of the cold box,  for minimizing heat losses and for personal protection against the risk of injuries by accidental contacts. For cold points, the insulation is based on synthetic rubber,
covered with aluminum sheets. Vacuum jacketed pipes are used for long-distance connections.

%% file: sections/shaft.tex
\subsection{Support structure in the shaft}
\label{sec:shaft}

\begin{figure}
\includegraphics[width=\columnwidth]{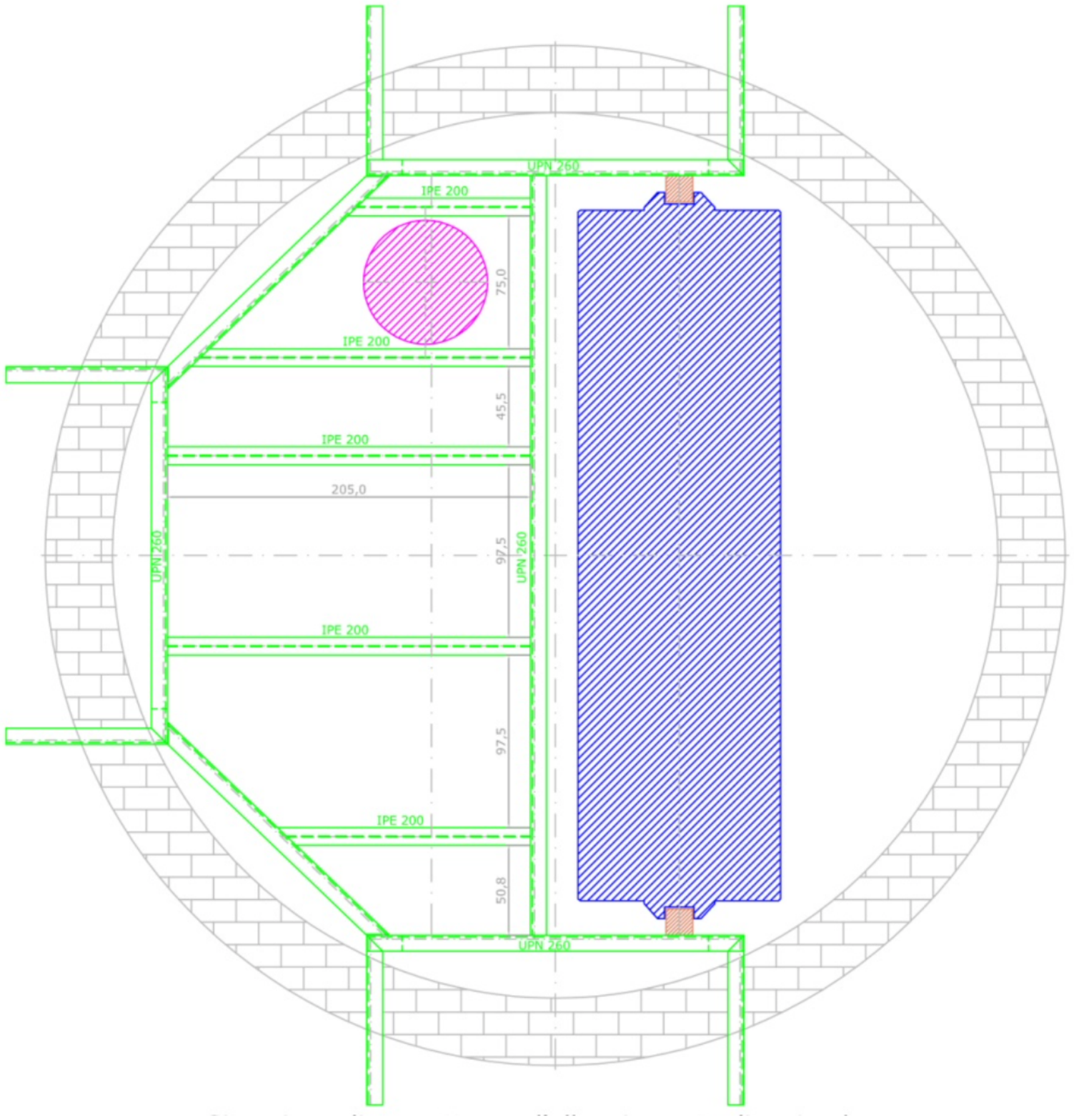}
\caption{Horizontal cut  of the mineshaft showing the stainless steel structure, in green, for  positioning  the   column, in magenta. The blue rectangle on the right is the elevator. }
\label{fig:design}
\end{figure}

\begin{figure}
\includegraphics[width=\columnwidth]{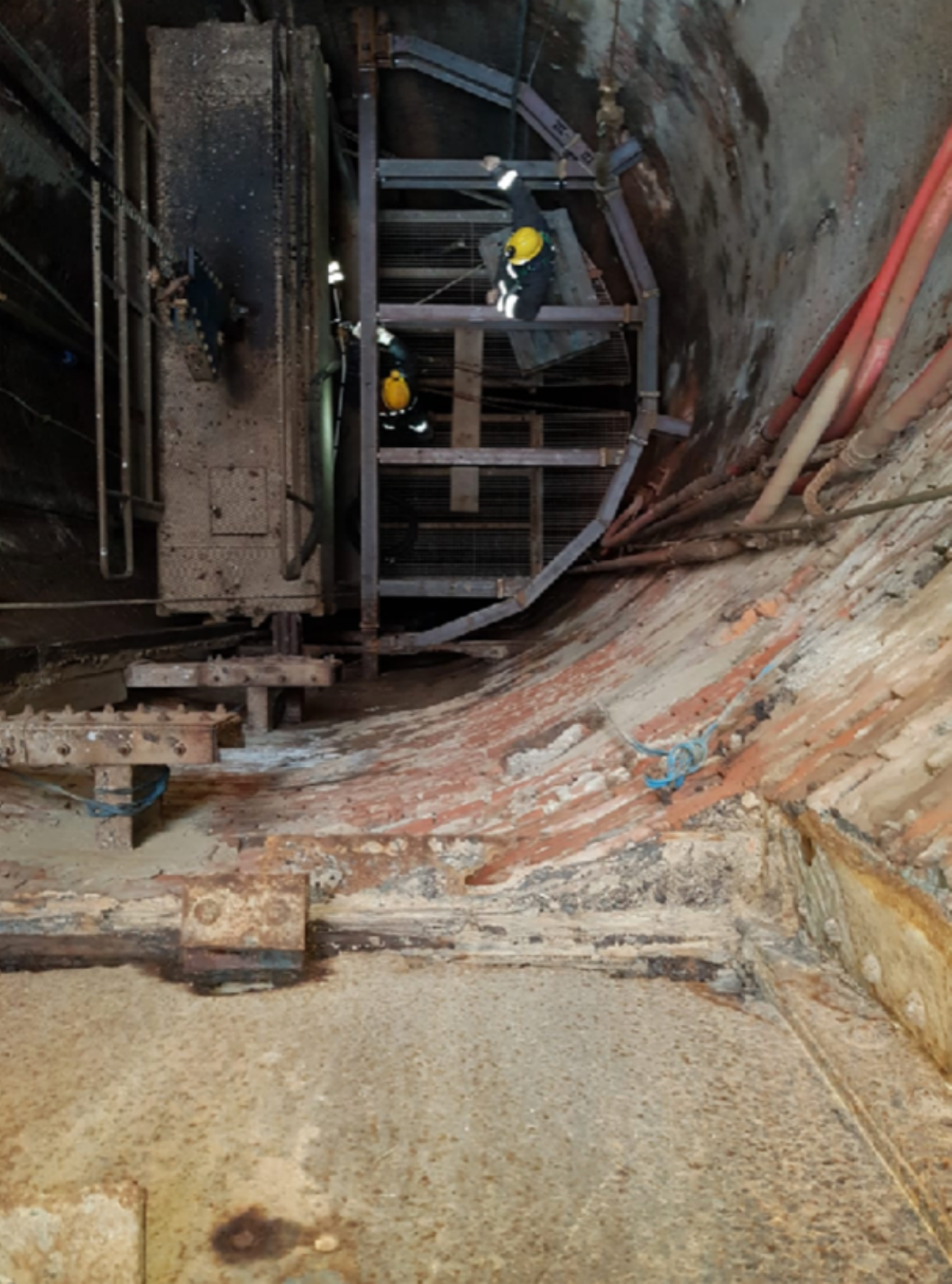}
\caption{Installation of the first support structure in the shaft of the Carbosulcis mine, Seruci site.}
\label{fig:palchetto}
\end{figure}

The support structure of the column is made  of aus\-ten\-itic steel and is assembled  by bolted connections. 
It is made of   discrete  structures, shown in \reffig{design}, spaced in the vertical direction by \AriaPlatformVerticalSpacing. 
To keep a safety margin, three supporting structures  per central module are foreseen, each one able, in principle, to hold the module independently.
The anchoring shelves penetrate the rock up to  an average depth of about \AriaDepthCentralSupportWall\ for the central support, and \AriaDepthTopBottomSupportWall\ for the other two. 
For filling the \SI{300}{\mm} openings,  a cement based thixo\-tropic  mortar is used,  with high mechanical strength and compensated shrinkage.
\reffig{palchetto} shows the installation of the first support structure  in the well.
Load tests were performed applying  a \AriaCentralModulesWeight\ load and no significant deflection was observed.

%% file: sections/leak_test.tex
\section{Vacuum leak tests of individual modules}
\label{sec:leak}
Leak detection is a critical step in the construction of \Aria, since  its functioning depends on a high cold box insulating vacuum and   the distillation process should  not be contaminated from  air.    
For that reason, by the way, the pressure of the process column and related lines is kept   above the atmospheric pressure. 
The leak-check procedure  has to be quite strict, in particular for those lines that will undergo thermal stresses. 
Indeed, the column and the service lines will go back and forth from room temperature down to liquid argon/nitrogen  temperatures    several times during their lifetime.

An upper limit of \AriaHeliumLeakCheckWeldPerModule\ was set on the leak rate for each leak check performed on single modules during testing, mainly on welds.
Each column segment  was tested twice. 
The first phase of tests took place at the manufacturing company site (Polaris Engineering), where the column and the service lines were fully tested, before wrapping them around with  MLI. 
The second phase of the leak checks, carried out at CERN, CH, included also a full check of the cold box  and bellows. For the tests, each module was closed temporarily with end-caps, the space between the cold box and the distillation column was evacuated with a turbopump system, and the column and the service lines were filled with a mixture of 90\% of air and 10\% of helium. In this way, the potential leak can  be found by the leak detector associated with the turbopump system. All the modules were validated in a two-step approach to confirm a  leak rate smaller than \AriaHeliumLeakCheckWeldPerModule\ on each module. 
Since there are \AriaModulesNumber\ column segments in total, the total leak rate is expected to be smaller than \AriaHeliumLeakCheckTotal\ at room temperature.
An additional one-off leak test was performed at CERN to validate module tightness after a thermal cycle down to \LArNormalTemperature. The reboiler unit was chosen for this test, due to its complex internal weld configuration, and tightness below \AriaHeliumLeakCheckWeldPerModule\ was again confirmed.



%% file: sections/prototype.tex
\section{Performance test at total reflux  with a prototype column.}
\label{sec:prototype}

To verify the theoretical calculations about the distillation performance, and test the mechanical and cryogenic infrastructure prior to  column installation in the mineshaft, a prototype plant was built in a surface building.

\subsection{Prototype construction}
\label{sec:prototype_construction}

\begin{figure}[ht!]
\centering
\includegraphics[width=\columnwidth]{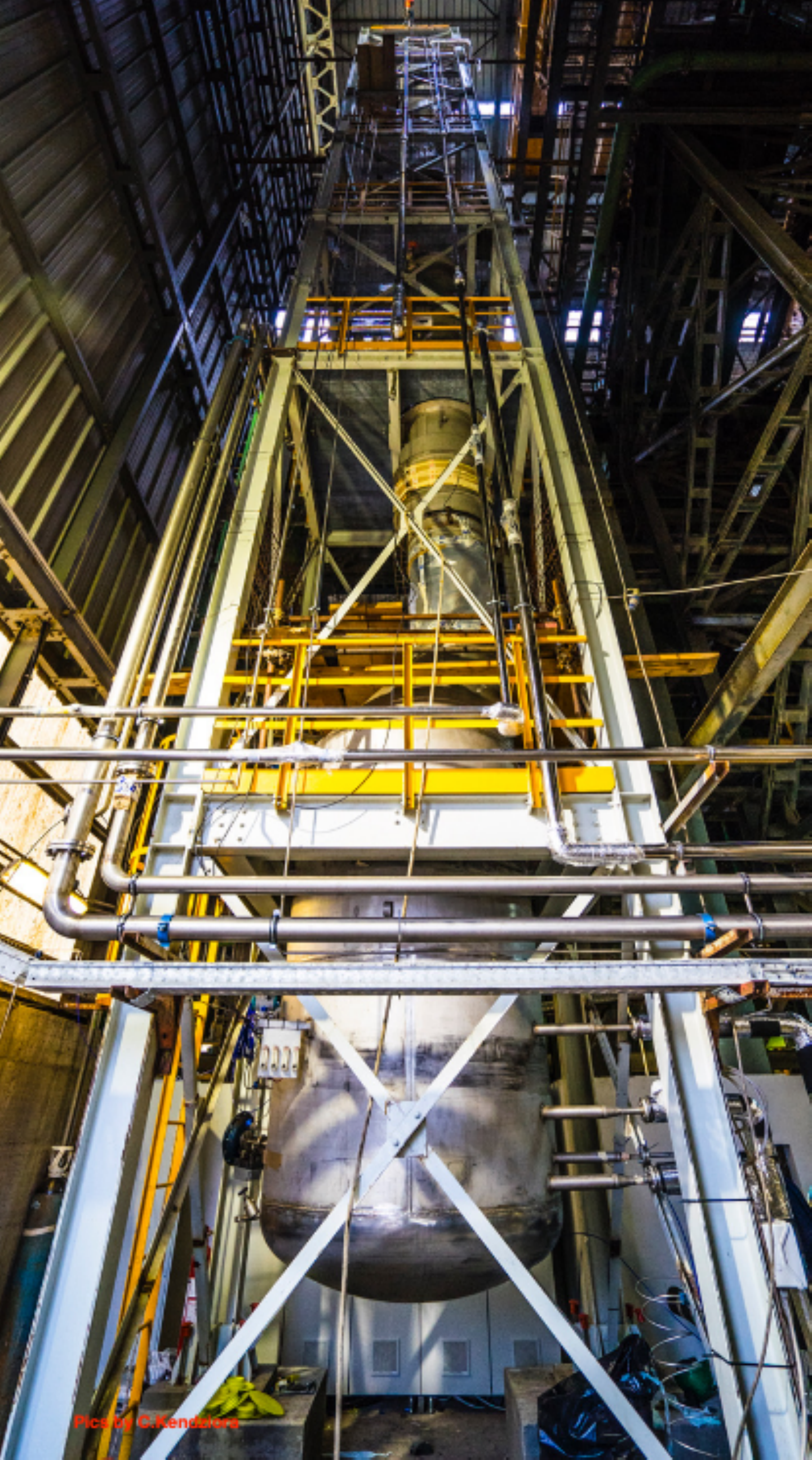}
\caption{The prototype \Aria\ plant in the Laveria building of the Carbosulcis mine, Nuraxi-Figus site, viewed from the basis of the column.}
\label{fig:plant}
\end{figure}

\begin{figure}[ht!]
\centering
\includegraphics[width=\columnwidth]{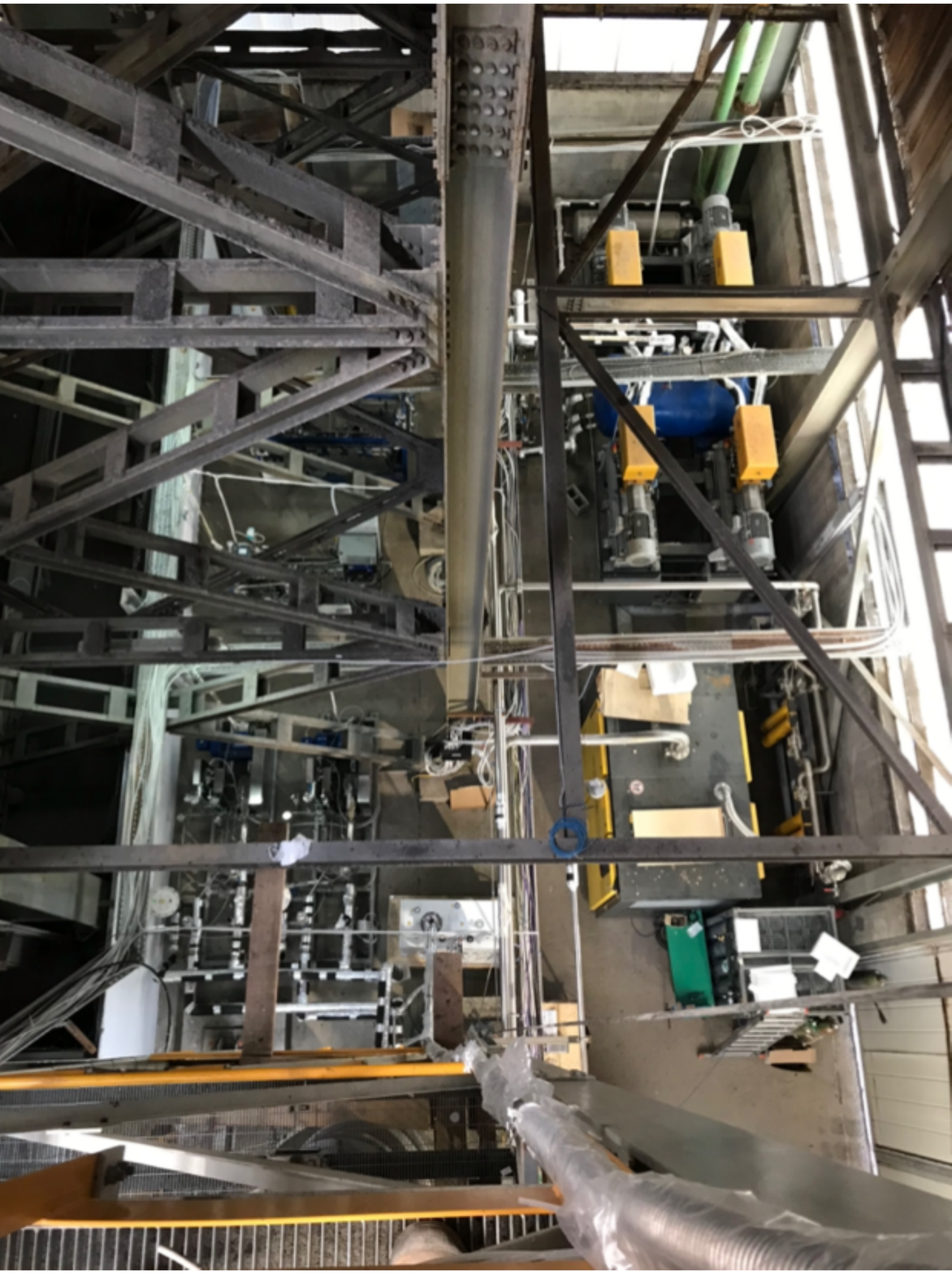}
\caption{Aerial view of the  prototype \Aria\ plant located in the Laveria building of the Carbosulcis mine, Nuraxi-Figus site. From bottom left, clockwise, the liquid pumps, the cryocoolers and the gas compressor.}
\label{fig:cryo}
\end{figure}

The prototype plant is a short version of the \Aria\ column using only the reboiler, the condenser and one central module, for a total height of \AriaSeruciZeroTotalHeight, together with the auxiliary equipment, which is the same as that of the full column. 
It is located in the Laveria building of the Carbosulcis mine, Nuraxi-Figus site, as shown in \reffig{plant} and \reffig{cryo}. 
The mechanical support, made of galvanized and cold-painted carbon steel, consists of a square base structure with four feet of concrete and a modular iron pillar structure equipped on each side with  two diagonal support beams.
The structure includes seven level platforms, to allow the presence of operators along the column height. Though self-supporting, for additional safety, the support is fixed to the building  structure at two different heights.

After welding together the three modules,   the column and the four service lines were leak checked  with a calibrated leak detector.  An external calibration leak was used to estimate the helium diffusion time along all the lines. This turned out to be between four and twenty seconds, depending on the line. Therefore it was  decided to wait at least two minutes between every leak check to make sure that a possible signal could be associated with the precise tested weld. The standard technique of filling sealed bags with helium around the welds was used for all the procedure. The helium bags, once filled with helium, were not removed until the last leak check. Using this method  an upper limit of \AriaHeliumLeakCheckWeldPerModule\ was set on all the welds between the modules.

Leak detection will become increasingly more difficult during the assembly of the modules in the shaft. With the leak test procedure just described, the increased size of the column, as the modules
are assembled together, will cause a much longer response time of the leak detection system, reducing its sensitivity. To overcome this difficulty, the use of some new tools is foreseen. Devices called clamshells, 
developed at CERN, will surround the welds and create a  small sealed space that can be quickly evacuated. Helium will flow inside the tube/column, and the potential leak in the weld can be therefore detected with a very fast response time.

\subsection{Prototype Operation}
\label{sec:prototype_operation}

For the commissioning of the prototype plant, different purity grades of \ce{N2} were used both in the auxiliary circuit for  cooling, and in the process circuit for the distillation inside the column.
The operating parameters of the auxiliary system were  about the same as those  discussed in  \refsec{plant}.

A dedicated slow-control system monitors and controls the distillation process and ll equipment in real time.
This system uses LabVIEW (NI) as system-design platform and development environment, and it is organized with 
a distributed layered architecture. The control cabinets are interconnected  by a private WLAN network, inside the CarboSulcis network, with a Real Time Controller (NI cRIO 9039)  reading out the data of the different expansion chassis (NI 9049) distributed over the network. In addition, 
 PRO\-FI\-BUS, a standard for fieldbus communication in automation technology, is  integrated into the  system  to  control  third party equipment such as compressors, vacuum gauges and vacuum pumps. The slow control also features advanced controls such as  Proportional Integral Derivative control, cascade control, threshold logic, interlocks over valves, inverters, and temperature controllers.  Historical data are stored in  a relational data base (PostgreSQL). 

Plant operation started with  feeding the cooling liquid nitrogen   to the auxiliary circuit from the external storage tank and nitrogen of purity grade \AriaSeruciZeroPurityGradeNitrogen\ into the column.
Eight hours
were needed to reach the target temperature.  The total amount of nitrogen  filling  the column was estimated by taking into account that it  was stored in  \AriaSeruciZeroNumGasBottles\ gas bottles  of \AriaSeruciZeroVolumeGasBottles\ each, with an initial pressure of  \AriaSeruciZeroInitialPressureGasBottles\ and a final pressure of \AriaSeruciZeroFinalPressureGasBottles. Using the Peng-\-Robin\-son equation of state, the total mass is then  \AriaSeruciZeroNitrogenTotalWeigth.

\begin{figure}
\includegraphics[width=\columnwidth]{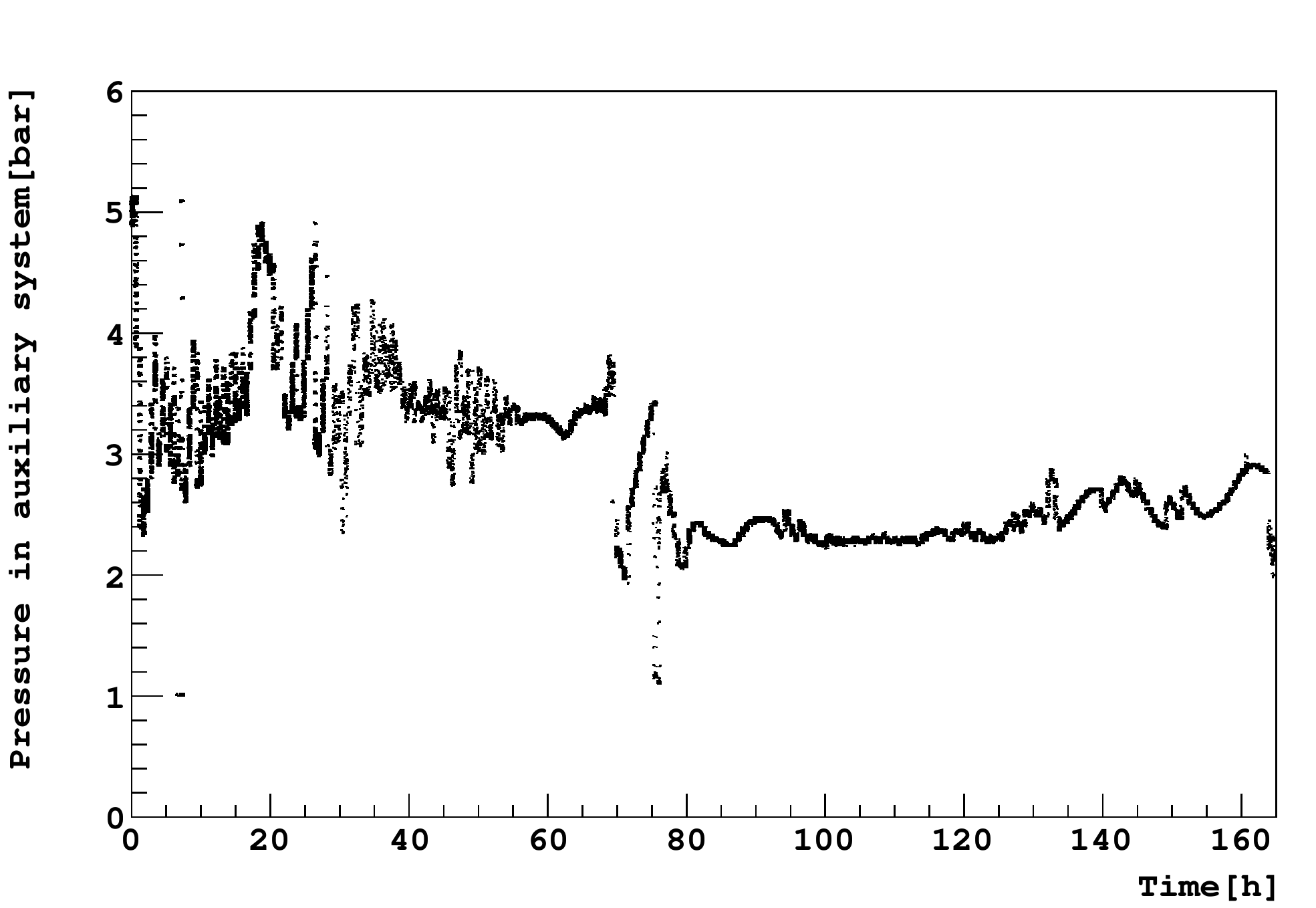}
\caption{Measured  pressure in the auxiliary system downstream of the compressor  vs time, for \ce{^29N_2}-\ce{^28N_2} distillation in the prototype plant.}
\label{fig:pressure_gas}
\end{figure}
\begin{figure}
\includegraphics[width=\columnwidth]{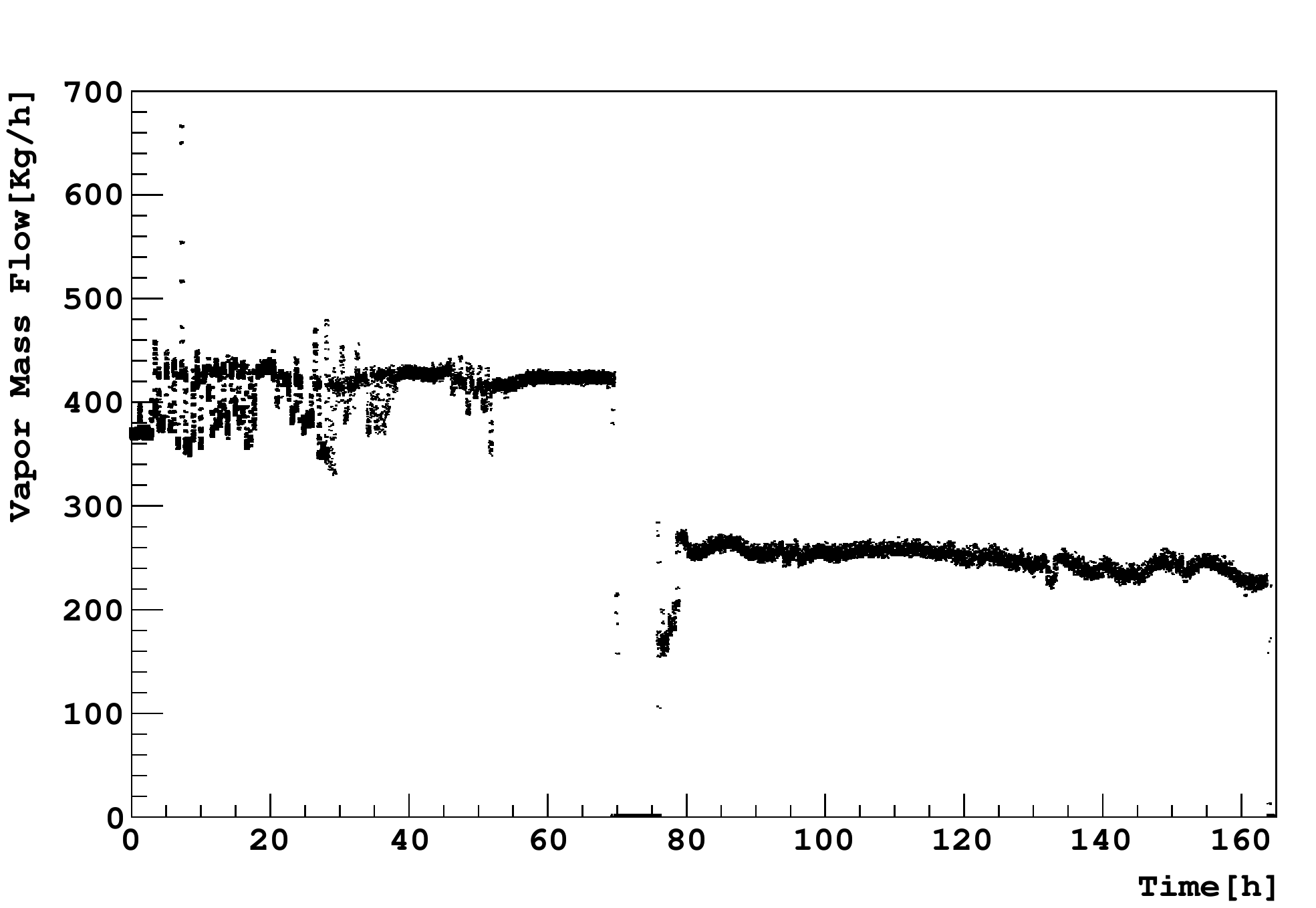}
\caption[]{Measured  vapor  mass flow in the auxiliary system downstream of the compressor vs time, for \ce{^29N_2}-\ce{^28N_2} distillation in the prototype plant.}
\label{fig:flow}
\end{figure}
\begin{figure}
\includegraphics[width=\columnwidth]{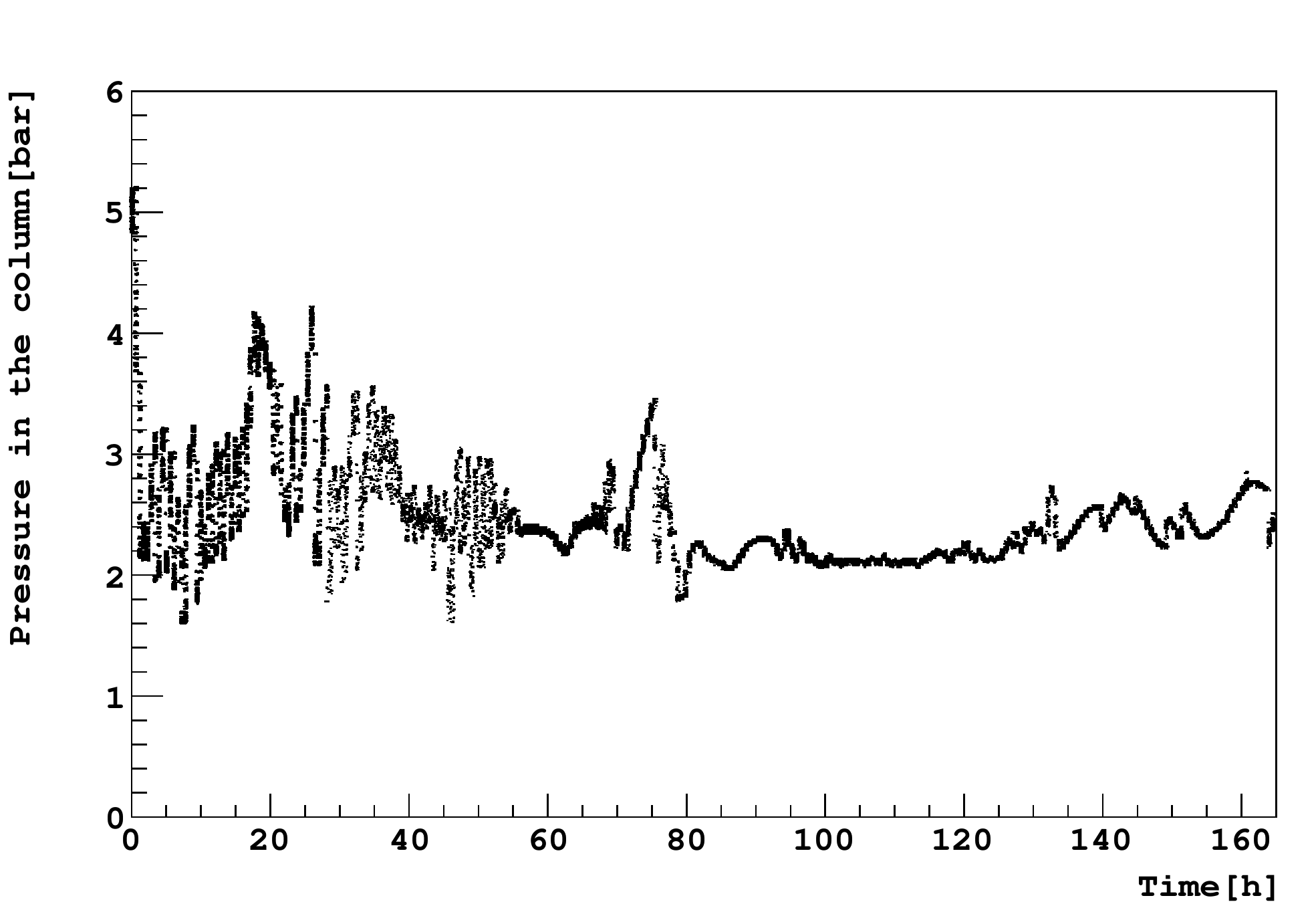}
\caption{Measured pressure inside the column in the top vs time, for \ce{^29N_2}-\ce{^28N_2} distillation in the prototype plant.}
\label{fig:pressure}
\end{figure}
\begin{figure}
\includegraphics[width=\columnwidth]{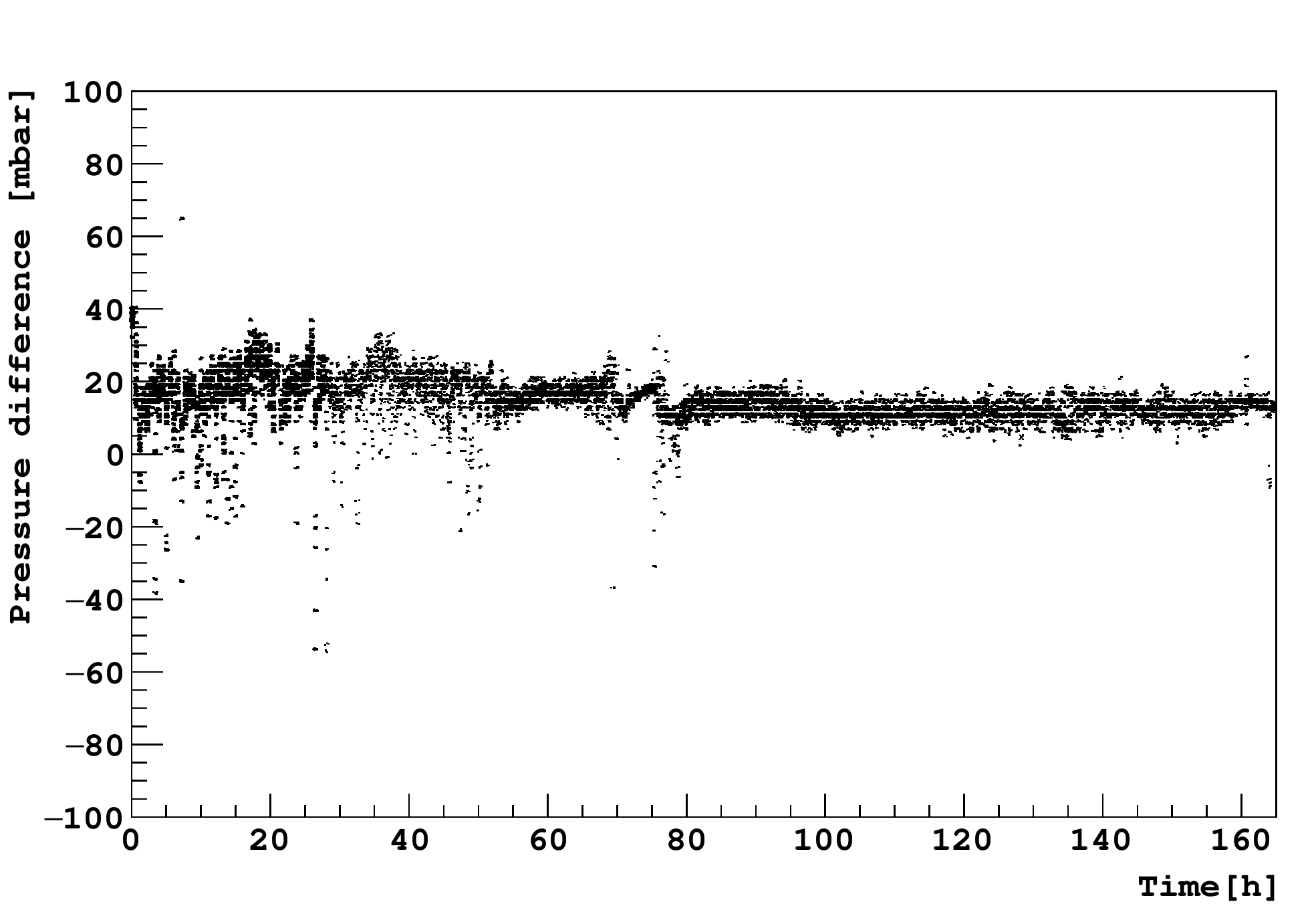}
\caption[]{Measured   pressure difference condenser vs. reboiler in the column vs time, for \ce{^29N_2}-\ce{^28N_2} distillation in the prototype plant.}
\label{fig:drop}
\end{figure}

The  measurements reported in this paper  refer to two distillation runs of the plant,  \AriaSeruciZeroNominalRunDuration\ in total,  with two different screw-rotary compressor (C3) settings, with the column operated at total reflux. The two runs started and stopped with switching on and off the compressor and, with some delay, the pumps.
 \reffiginitpar{pressure_gas} and \reffig{flow}  show the measured   pressure vs. time and mass  flow  vs. time in the auxiliary system, downstream of the compressor. 
For these first two runs, an automated feedback system, foreseen in the plant design, regulating the  flow downstream of the compressor, was not used. The auxiliary system gas pressure  and flow stability were  guaranteed only  regulating by hand  a bypass valve between the compressor and the gas flow meter. A better stability was reached during the second run, where fluctuations were limited to   \AriaSeruciZeroPressureStability\ and  \AriaSeruciZeroFlowStability, as shown in \reffig{pressure_gas}  and \reffig{flow}, respectively.
The pressure inside the column was measured by digital pressure transmitters with diaphragm seal measuring cell,
located respectively below the first distributor from the top and right above  the reboiler.
  \reffiginitpar{pressure} shows the measured   pressure inside the column in the top vs. time. The different pressure in the column compared to what is expected for argon, as discussed in \refsec{plant}, comes from the different thermodynamic properties of the nitrogen and the operating temperature gradients of the heat  exchangers of the reboiler and of the condenser of about \AriaSeruciZeroHeatExchangerTemperatureDifference.
Since nitrogen was used both for  cooling and as distillation fluid,  the   mass flow rate in the cooling circuit was the same as that inside the column. From \reffig{flow} it can be deduced, then, that the  mass flowrates during this test ranged between  the maximum allowed flow in the column by the packing producer, as discussed in  \refsec{column}, and half of it. 
The pressure drop along the prototype column during the second run was  of the order of \AriaSeruciZeroPressureDrop, as shown in \reffig{drop}. 
 Therefore, for the following calculations, it is  possible to assume that both pressure and temperature are constant along the column.
 The nitrogen temperature  inside the column was  derived from the pressure measurement using the Antoine equation \cite{doi:10.1021/je60033a014}.
From the data of \reffig{pressure}  it follows that, during the second run, the temperature ranged from \AriaSeruciZeroLowT\ to \AriaSeruciZeroHighT.
The measured vacuum level in the cold box during the two runs was  stable around  \AriaSeruciZeroMeasuredVacuum.


\subsection{Expected values for nitrogen distillation}
\label{sec:prototype_nitrogen}

The nitrogen molecule, \ce{N_2}, is mainly formed by  two stable  isotopes, \ce{^14N} and \ce{^15N}, leading to  an isotopic fraction of  \NitrogenFourteenIsotopicFraction\ for the  \ce{^28N_2}  and \NitrogenFiftteenIsotopicFraction\ for the  \ce{^29N_2}, and,  therefore, to  an isotopic  ratio, $R_{\ce{N_2}}$, between the two molecules, of  
\NitrogenIsotpicRatio.
 The relative volatility between \ce{^29N_2} and \ce{^28N_2}, $\alpha_{28-29}$,  is given, according to \cite{ANDREEV2007247}, by 
$\ln\alpha_{28-29}$ = 0.846/T - $6.9 \times 10^{-3}$, where T is the temperature in Kelvin, implying   $\alpha_{28-29}$=\AriaSeruciZeroAssumedAlpha, at the mean column operating temperature of \AriaSeruciZeroMeanT.
This value of the relative volatility is large enough to give a sizeable  separation, at total reflux,  even with the prototype column that nominally has only about \AriaSeruciZeroNumberOfTheorecticalStages\ theoretical stages, namely of  $S^0_{28-29}$=\AriaSeruciZeroTotalRefluxCalculatedSeparation. 

\subsection{Distillation measurements}
\label{sec:prototype_results}

\begin{figure}
\centering
\includegraphics[width=\columnwidth]{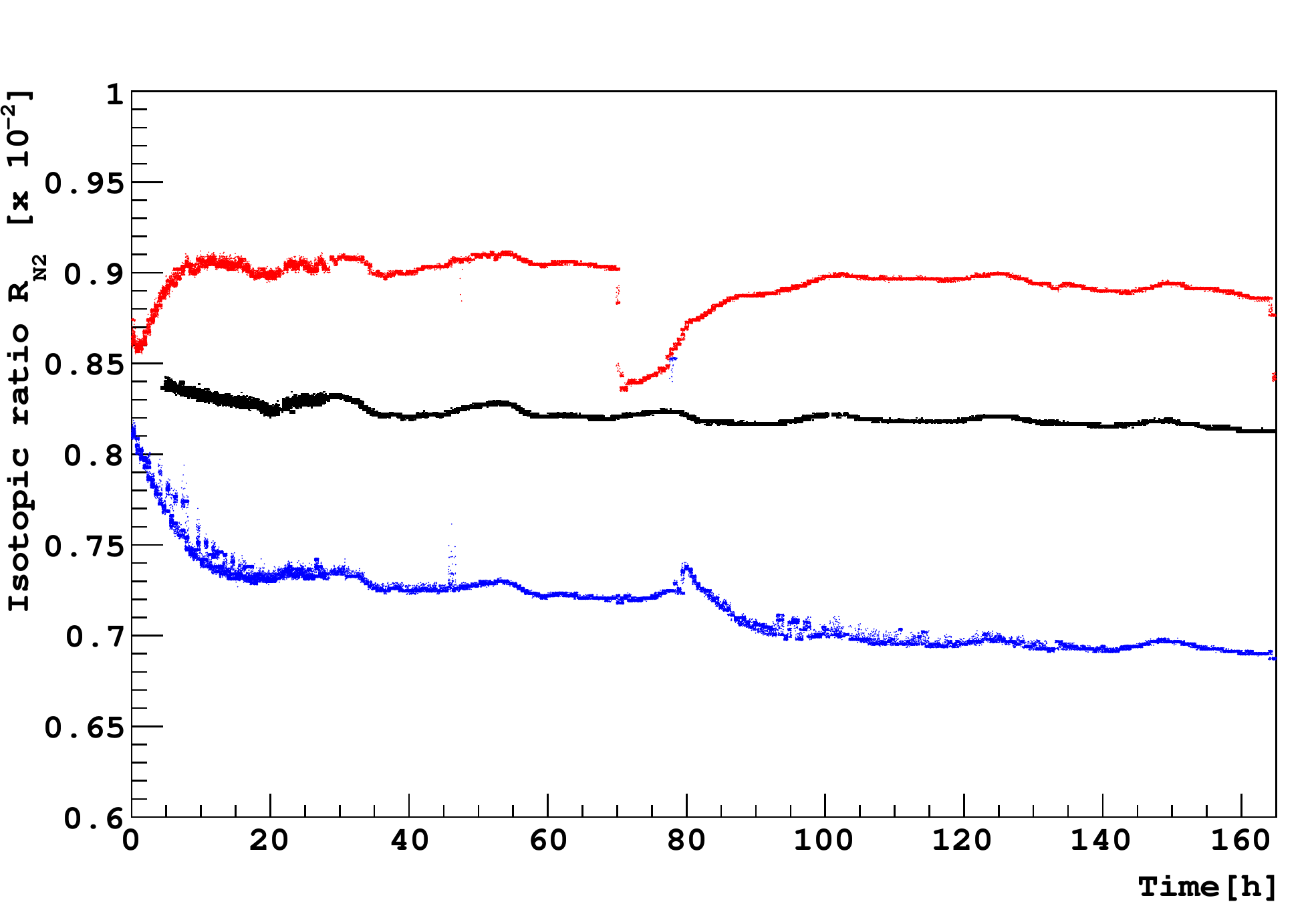}
\caption[]{Reboiler (red), condenser (blue), and feed (black), isotopic ratio  $R_{\ce{N_2}}$ vs. time  for \ce{^29N_2}-\ce{^28N_2} distillation in the prototype plant. }.
\label{fig:rcf}
\end{figure}

\begin{figure}
\centering
\includegraphics[width=\columnwidth]{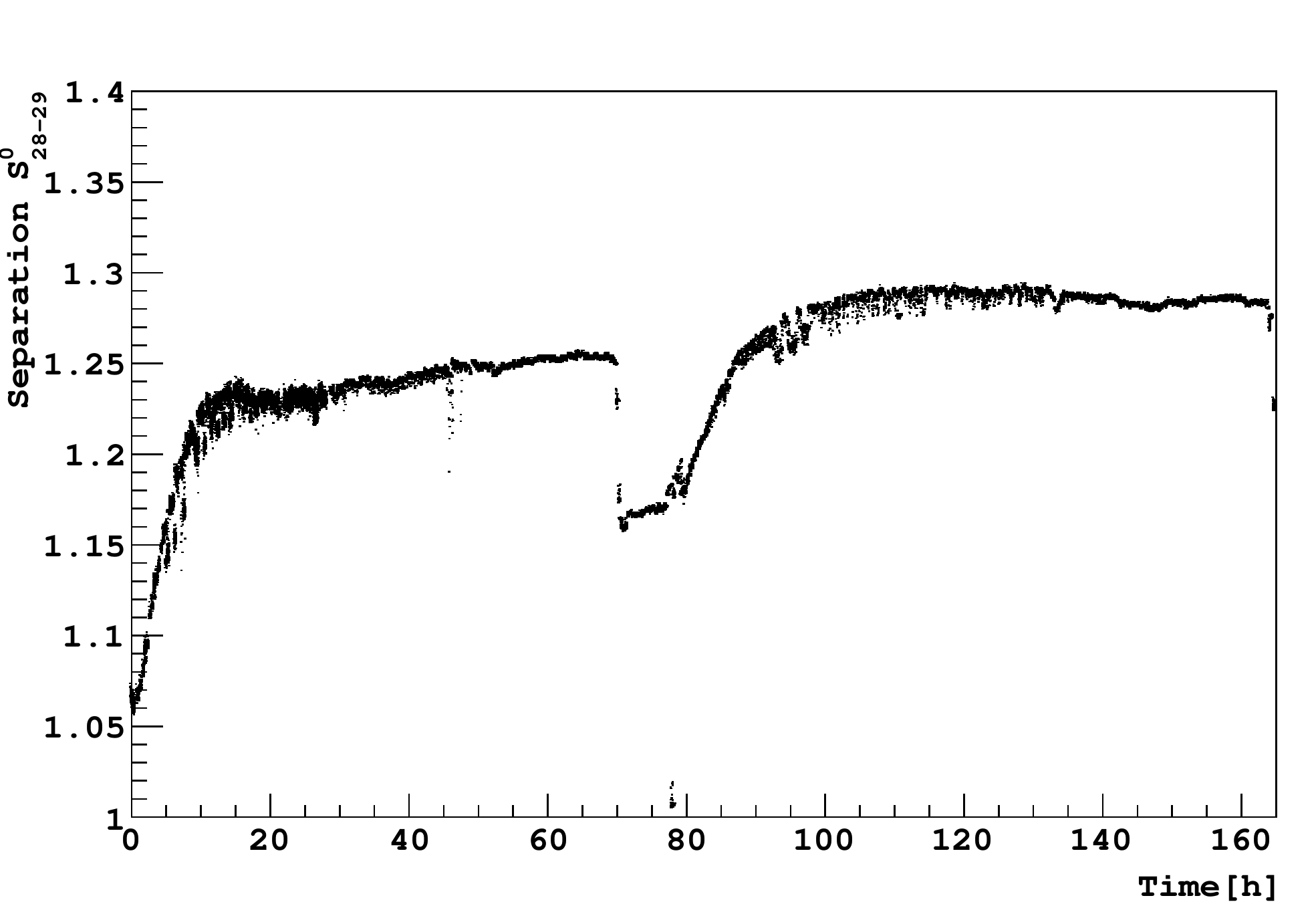}
\caption[]{Separation factor $S^0_{28-29}$  for \ce{^29N_2}-\ce{^28N_2} distillation in the prototype plant. }
\label{fig:separation}
\end{figure}

A quadrupole mass spectrometer (Extrel MAX-300) measured the fluid composition, sampling in the reboiler, in the condenser, and in the feed line at the output of the gas bottles, using up to \AriaSeruciZeroCapillaryLength\ long and \AriaSeruciZeroCapillaryDiameter\ diameter copper capillaries.
With this mass spectrometer, the peaks corresponding to \ce{^28N_2} and \ce{^29N_2} are well separated, 
and, therefore, isotopic ratio measurements were directly  taken from the peak height ratio.
In \reffig{rcf} the measured isotopic ratios $R_{\ce{N_2}}$ vs. time from the reboiler, condenser and feed are shown.
When the plant started operation, the three isotopic ratios were the same. With time,  they started to diverge, as expected with the  distillation taking place, and, after some time,   they reached a  plateau value. It should be noticed that at the end of the first run, the isotopic ratio in the reboiler dropped almost to the feed value while the one of the condenser increased only after about 10 h. This is due to the fact that when the compressor and the pumps are  switched off, i.e. the distillation process is stopped, the liquid present in the columns sinks quickly in the reboiler under gravity, and mixes with that already present there,    whereas this is not the case for the vapor.   
From \reffig{rcf} it can also be noticed that  the  isotopic ratio  of the feed gas  is not exactly equal to the natural isotopic ratio value  discussed in \refsec{prototype_nitrogen} and this is due to  the mass spectrometer  not fully calibrated before use. Also the isotopic ratio values slightly drifted over time, probably due to some internal component of the spectrometer becoming dirty. Anyhow, the separation, $S^0_{28-29}$, defined   as $R_{\ce{N_2}}$(con\-den\-ser)/$R_{\ce{N_2}}$(re\-boiler), depending only on the ratio of the two $R_{\ce{N_2}}$, it is not affected by this drift to first order.
\reffiginitpar{separation} shows the separation, $S^0_{28-29}$, vs time.  
\subsection{Measurement interpretation}
\label{sec:prototype_results_intepretation}

\begin{figure}
\centering
\includegraphics[width=\columnwidth]{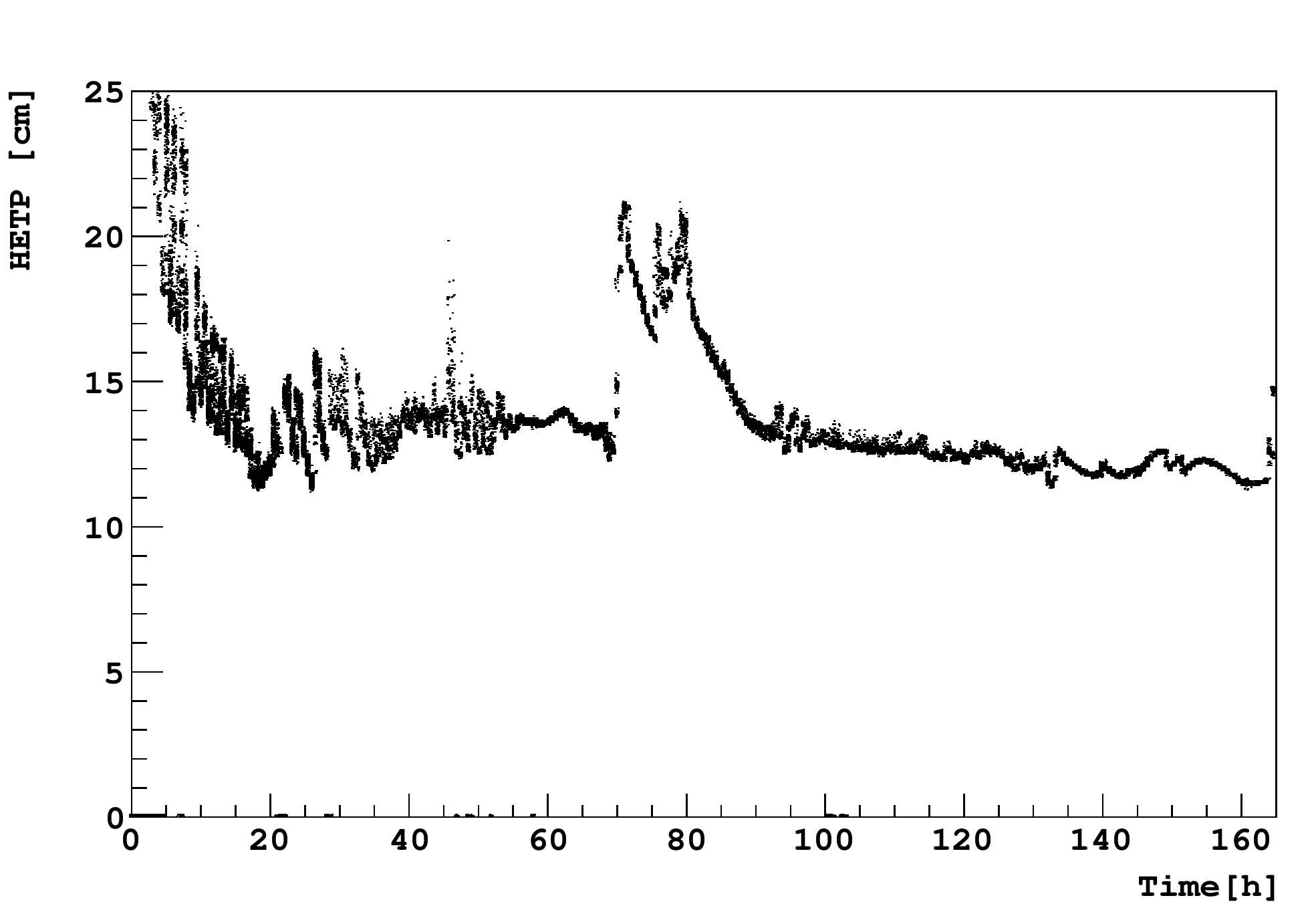}
\caption[]{HETP vs. time for \ce{^29N_2}-\ce{^28N_2} distillation in the prototype plant. }
\label{fig:c_NumStadi_I_vs_date_KaeTotal}
\end{figure}

From the measured separation $S^0_{28-29}$ and the calculated relationship between $\alpha_{28-29}$ and temperature, as discussed in \refsec{prototype_nitrogen}, it is possible to derive the effective number of theoretical stages and, knowing   the packing height per module of \AriaColumnModulePackingHeight, as  discussed in \refsec{column}, the  HETP vs. time,  as shown in \reffig{c_NumStadi_I_vs_date_KaeTotal}. The best value obtained during the two runs is  about \AriaSeruciZeroEffectiveHETP.
This value of HETP is in broad agreement with the assumed value  for  argon of \refsec{performance}. This agreement represents a validation of  the concept of  cryogenic distillation with this plant.
However, it should be underlined that the tests described here were performed, in terms of   pressure inside the column, outside of the range for which the vendor provides comparison data, and at a value of the sizing parameter $F$ of \AriaSeruciZeroFValueNitrogen, which is different from that calculated for argon in \refsec{column}. 
 The observed transient, i.e. the time needed to reach plateau operation  in \reffig{separation}, 
turns out to be about \AriaSeruciZeroTimeToReachStableConditionsSimulation. 
It is important to point out that the time to reach the steady state  is strongly dependent on the fluid to be distilled, the duty at the reboiler, and the number of theoretical stages. Further investigation is therefore required before extrapolating this value to the  \Aria\  column performance with argon.


%% file: sections/performance.tex
\section{Projected performance of \Aria\ with argon, at finite reflux}
\label{sec:performance}

\begin{figure*}
\centering
\includegraphics[width=0.75\textwidth]{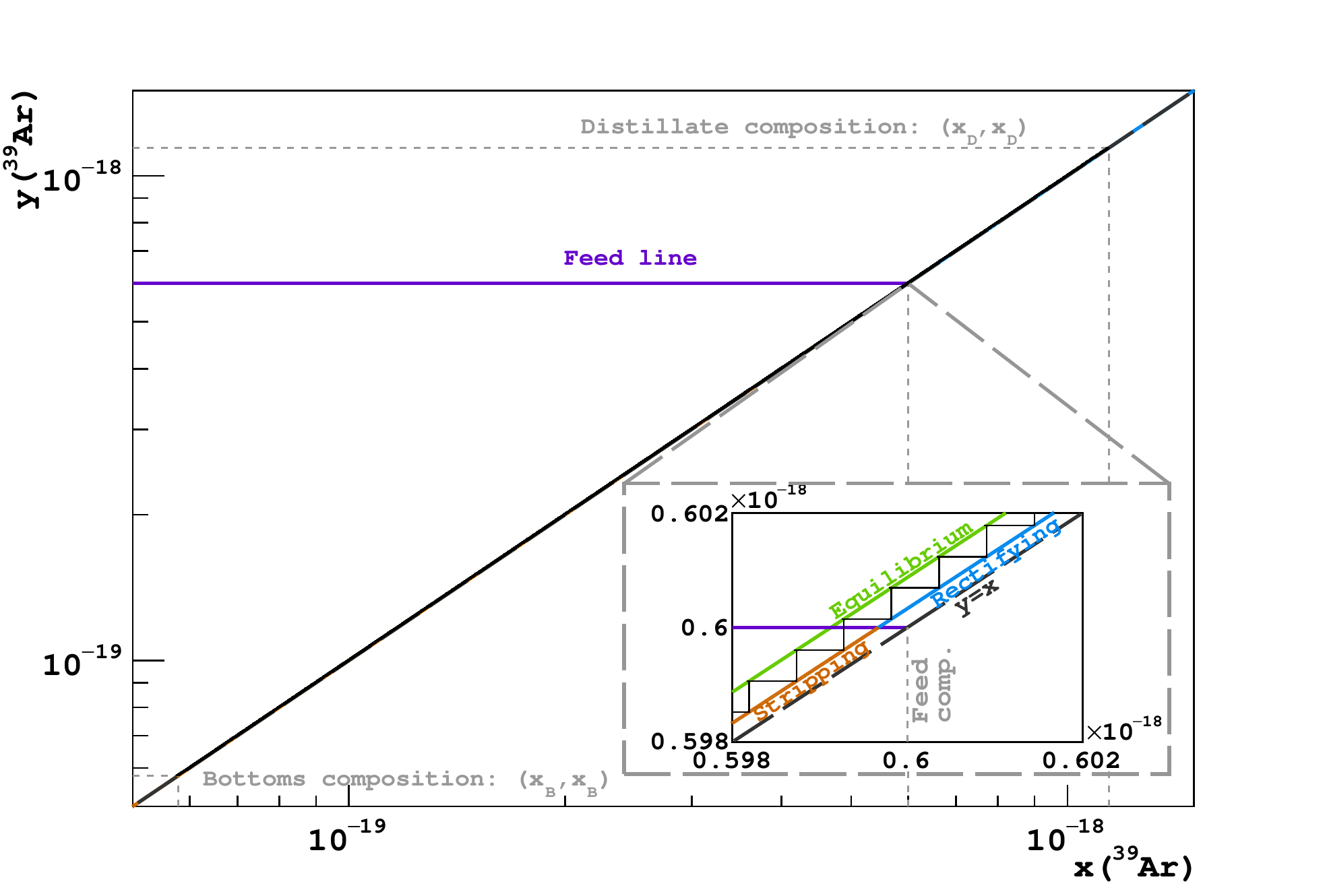}
\caption{McCabe-Thiele diagram for the \ce{^39Ar}-\ce{^40Ar} distillation with the input parameters of \reftab{input_parameters}, for $B/F$= \AriaNominalHeavyRecovery: mole fraction of \ce{39Ar} in the liquid  phase, $y(\ce{39Ar})$, vs. mole fraction of \ce{39Ar} in the vapor phase, $x(\ce{39Ar})$. The insert is a 
blow-up of the region indicated by the shaded lines.}
\label{fig:mccabe}
\end{figure*}

The McCabe-Thiele method~\cite{McCabe:1925be} was used to calculate the performance of argon distillation in \Aria\ with finite reflux.
This method was already applied to cryogenic distillation by collaborations using xenon as  active target for dark matter search \cite{Wang2014,ABE2009290,xenon}, but it is fair to say that it was never validated with argon.
The  input parameters  of the calculation are summarized in
\reftab{input_parameters} and it was also assumed the feed to be a  saturated vapor.   The relative  volatility, $\alpha_{39-40}$, is assumed constant along the column height and equal to the value corresponding to the mean operating temperature of the column. 
\begin{SCtable}
\centering
\caption[]{Input parameters of the calculation of \ce{^39Ar}-\ce{^40Ar} distillation with the McCabe-Thiele method. $\rho_L$ is the liquid  argon density at equilibrium   at \AriaMeanOperatingTemperature\ and $x_F$   the  molar fraction of \ce{^39Ar} in the feed.  The other parameters are described in the text.}
\label{tab:input_parameters}
\begin{tabular}{ll}
\hline\noalign{\smallskip}

	parameter		& value \\
	\noalign{\smallskip}\hline\noalign{\smallskip}
	$x_F$ & \AArArThreeNineOverArFourZeroRatioUAr\ \\
	$x_B$ & \AArArThreeNineOverArFourZeroRatioUArDividedByTen\ \\
	$\alpha_{39-40}$ & \AriaArVolatiityRatioMean\ \\
	$\rho_L$ & \AriaLiquidDensity \\
	$\rho_V$ & \AriaVaporDensity \\
	$d$ & \AriaColumnInnerDiameter \\
	$N$ & \AriaNumberOfTheorecticalStages \\
	     $\hat{V}_L$ & \AriaLiquidFlowForTheoreticalPlate \\
\noalign{\smallskip}\hline
\end{tabular}
\end{SCtable}
\begin{figure}
\includegraphics[width=\columnwidth]{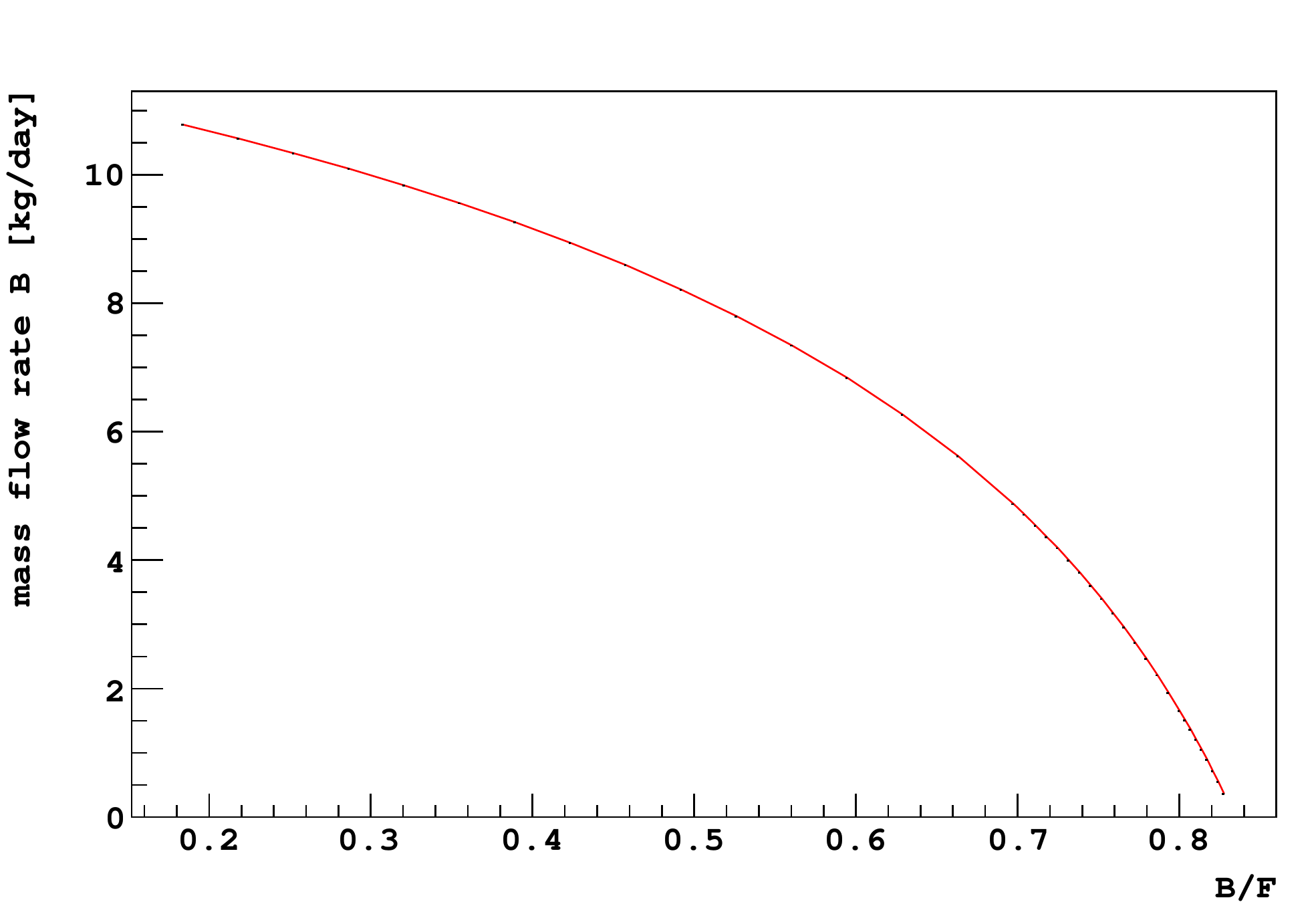}
\caption[]{\ce{^39Ar}-\ce{^40Ar} distillation with the McCabe-Thiele method: $B$ mass flow vs.  $B/F$.}
\label{fig:rates}
\end{figure}
The calculation was performed for individual  values of  $B/F$, where $B$ and $F$ are the mass flow rates in the bottom and feed streams, respectively.
\reffig{rates} shows  $B$  vs.  $B/F$. We consider as benchmark value $B/F$ to be \AriaNominalHeavyRecovery, a reasonable assumption given the UAr is a valuable  material.
The McCabe-Thiele diagram corresponding to  this benchmark value  is shown in \reffig{mccabe}.
The output parameters of the calculation 
are shown  in  \reftab{rates}.  The calculation also yields the location of the feed point in the column, which  turns out to be at the top of the seventh module, i.e. at about 20\% height from the top of the column. This is where the feed  connections are located.
The obtained value of  $S_{39-40}$, the separation of \refeqn{Fenske2} calculated at finite reflux, can be compared with that obtained at total reflux,  $S^0_{39-40}$, which is   \AriaDepletionPerPassTotalReflux.
If $x_B$ were required to be \AArArThreeNineOverArFourZeroRatioUArDividedByTwenty, then $B$ would become \AriaBottomStreamForNominalHeavyRecoveryAndFactorTwenty, with the same feed point.
\begin{SCtable}
\centering
\caption[]{Output parameters of the calculation of \ce{^39Ar}-\ce{^40Ar} distillation with the McCabe-Thiele method, for $B/F$= \AriaNominalHeavyRecovery.  The various parameters are described in the text.
}
\label{tab:rates}
\begin{tabular}{ll}
\hline\noalign{\smallskip}

	parameter		& value \\
	\noalign{\smallskip}\hline\noalign{\smallskip}
	$B$ &  \AriaBottomStreamForNominalHeavyRecovery\ \\
$F$ &  \AriaFeedStreamForNominalHeavyRecovery\ \\
$R$ & \AriaRefluxRatioForNominalHeavyRecovery\ \\ 
$x_D$ & \AArArThreeNineOverArFourZeroRatioUArDistillate\ \\

 $S_{39-40}$ &\AriaNominalSeparation\ \\
\noalign{\smallskip}\hline
\end{tabular}

\end{SCtable}

The dominant systematic uncertainties in this calculation come from  the uncertainty on the mean $\alpha_{39-40}$  value   and on the number of theoretical stages, $N$. As discussed in  \refsec{requirements}, $\alpha_{39-40}$ was  calculated from the measured value of $\alpha_{36-40}$. Another publication \cite{Boato:1960a} reports a lower value of $\alpha_{36-40}$ than that reported in  \refsec{requirements}, that would lead to $\alpha_{39-40}$=1.0013 at the mean operating temperature of the column of  \AriaMeanOperatingTemperature. At this value one would have a decrease in B by about 25\%. 
A 10\% variation on $N$ leads instead to a 30\% change in $B$. Were the measurements of HETP reported in \refsec{prototype}  confirmed in a run with argon, a  decrease in $B$ of this magnitude, compared to that of \reftab{rates},  could be expected. Eventually, all the output rates are proportional to $\hat{V}_L$, i.e. halving this value  leads to halving $B$.
 The effect  of varying  $\alpha_{39-40}$ along the column height according to the temperature profile, was estimated modifying the  standard   McCabe-Thiele calculation, with the equilibrium curve  assumed to be varying  stage by stage. A marginal difference in the final result was obtained.

A major assumption in the  above calculation  is the  binary distillation hypothesis, i.e. that  isotopes present in the gas other than \ce{^39Ar} and \ce{^40Ar} do not influence the calculation.
It is well known that \ce{^36Ar} and \ce{^38Ar}  have  sizeable isotopic fractions in \AAr, of 0.33\% and 0.06\%, respectively, though it has been reported  that their isotopic fraction is about forty times lower in \UAr\ \cite{Saldanha:2019zpu}. 
However, the assumption of a binary mixture is considered to be reasonable, for two main reasons.
On one hand, the two additional isotopes  are mostly  recovered in the distillate stream, because  their relative volatility to  \ce{^40Ar} is larger than one, and therefore  we expect  no significant difference  in the composition of the bottom stream.
On the other hand, the isotopic fraction of  both the distillate and the bottom flow of \ce{^36Ar} and \ce{^38Ar}  are anyway  expected to change only by a small factor  since, for each isotope $i$, 

\begin{equation}
\frac{B \cdot (x_B)_i}{F \cdot (x_F)_i} < 1 \;\;\; \mbox{and}\;\;\;  \frac{D \cdot (x_D)_i}{F \cdot (x_F)_i} < 1
\end{equation}

or 

\begin{equation}
(x_B)_i < 1.8 \cdot (x_F)_i  \;\;\; \mbox{and} \;\;\;  (x_D)_i < 2.2 \cdot (x_F)_i,
\end{equation} 
Therefore, the thermodynamic properties of the isotope mixture are  marginally changed during the distillation process, and so does
$\alpha_{39-40}$. 
 
Moreover, the presence of \ce{^36Ar} with a significant isotopic fraction can be very useful for the plant commissioning \emph{e.g.} with atmospheric argon, by measuring isotopic fractions along the column with a mass spectrometer. At total reflux, the separation factor $S_{36-40}$ is \AriaThirtySixDepletionPerPassTotalReflux. At finite reflux, a calculation with the McCabe-Thiele method with the same parameters as above, requiring a reduction factor of \AriaThirtySixDepletionPerPass, gives the results of \reftab{thirtysix}.

\begin{SCtable}
\centering
\caption[]{McCabe-Thiele method: output parameters for \ce{^36Ar}-\ce{^40Ar} distillation in a run with atmospheric argon. Feed, $F$, and bottom, $B$, mass flowrates of the feed  argon,  and molar fraction of   \ce{^36Ar} in the top distillate, $(x_D)_{36}$. The calculation was performed requiring the \ce{^36Ar} isotopic fraction to be reduced   by a factor of  \AriaThirtySixDepletionPerPass.}
\label{tab:thirtysix}
\begin{tabular}{ll}
\hline\noalign{\smallskip}

	parameter		& value \\
	\noalign{\smallskip}\hline\noalign{\smallskip}
	$B$ &  \AriaThirtySixBottomStreamForNominalHeavyRecovery\ \\
$F$ &  \AriaThirtySixFeedStreamForNominalHeavyRecovery\ \\
$(x_D)_{36}$ & \AArArThreeSixOverArFourZeroRatioUArDistillate\ \\
\noalign{\smallskip}\hline
\end{tabular}

\end{SCtable}

Another factor that has to be taken into account, when calculating the plant performance in terms of \ce{^39Ar} suppression, is the cosmogenic activation of the argon. Cosmogenic activation  occurs at the extraction site in Colorado, during transportation, and at the \Aria\ site, during the operation of the plant, since the argon to be processed is  stored on the surface. Preliminary studies  indicate that the dominant component of the activation comes from cosmic ray neutrons during  storage at the Aria site, and is of the order of \UArActivationFractionOfTargetRadioactivityInOnePass\ of the underground argon radioactivity after distillation, for a  \AriaUArForLowMassTotalProductionTime\ long run to obtain  \UArForLowMass\ of argon with a reduced isotopic fraction of \AriaDepletionPerPass.  

The \ce{^39Ar}  isotopic fraction is so low that it cannot be detected with a mass spectrometer. Therefore, to verify the performance of \Aria\ in terms of isotopic distillation,  a new experiment, \dia{} \cite{Aalseth:2020nwt}, based on  a radioactivity measurement, was recently designed and approved at the Canfranc Underground Laboratory (LSC), Spain. 
The experiment 
 is expected  to set an upper limit on the $DF$, at  90\% C.L.,  of \DartUpperLimitOnDepletionFactor. Therefore, it is expected to measure the residual \ce{^39Ar} content after distillation in the  commissioning phase of the \Aria\ plant with atmospheric argon with good precision.

%% file: sections/Conclusion.tex
\section{Conclusion and outlook}

The design, construction, prototype tests, and performance simulations of the Aria cryogenic distillation column that is currently in the installation  phase at  Carbosulcis S.p.A., in Nuraxi-Figus (SU), Italy were discussed in detail. 
The measurement with the prototype of a HETP in broad agreement with the expectations, validated the concept of the cryogenic distillation with this plant.
The successful run of the \Aria\ plant is expected to have a tremendous impact in the field of isotopic separation, with applications ranging from nuclear physics to medicine and beyond.

%% file: sections/Acknoledgment.tex
\section*{Acknowledgements}

The second phase of the leak checks, carried out at CERN, was performed under service agreement KN3155/TE.
We acknowledge the   professional contribution of  the Mine and  Electrical Maintenance staff of Carbosulcis S.p.A.
Part of the project funding comes from  {\em  Intervento finanziato con ri\-sor\-se FSC 2014-2020 - Patto per lo Sviluppo della Regione Sar\-degna}.
This paper is based upon work supported by the U. S. National Science Foundation (NSF) (Grants No. PHY-0919363, No. PHY-1004054, No. PHY-1004072, No. PHY-1242585, No. PHY-1314483, No. PHY- 1314507, associated collaborative grants, No. PHY-1211308, No. PHY-1314501, No. PHY-145\-5351 and No. PHY-1606912, as well as Major Research Instrumentation Grant No. MRI-1429544), the Italian Istituto Nazionale di Fisica Nucleare (Grants from Italian Ministero dell'Istruzione, Universit\`a, e Ricerca {\em ARIA e la Ricerca della
Materia Oscura} - Fondo Integrativo Speciale per la Ricerca (FISR) and Progetto Premiale 2013 and Commissione Scientifica Nazionale II). We acknowledge the financial support by LabEx UnivEarthS (ANR-10-LABX-0023 and ANR-18-IDEX-0001), the Natural Sciences and Engineering Research Council of Canada, SNOLAB, Arthur B. McDonald Canadian Astroparticle Physics Research Institute, and  the S\~ao Paulo Research Foundation (Grant No. \- FAPESP - 2017/26238-4). The authors were also supported by the {\em Unidad de Excelencia Mar\'{\i}a de Maeztu: CIEMAT - F\'{\i}sica de part\'{\i}culas} (Grant No. MDM 2015-0509), the Polish National Science Centre (Grant No. UMO-2019/33/B/\-ST2/\-02884), the Foundation for Polish Science (Grant No. TEAM/2016 - 2/17), the International Research Agenda Program AstroCeNT \- (Grant No.\- MAB/\-2018/\-7) funded by the Foundation for Polish Science from the European Regional Development Fund, the European Union’s Horizon 2020 research and innovation program under grant agreement No 962480, the Science and Technology Facilities Council, part of the United Kingdom Research and Innovation, and The Royal Society (United Kingdom). I.F.M.A is supported in part by Conselho Nacional de Desenvolvimento Científico e Tecnol\'ogico (CNPq).  We also wish to acknowledge the support from Pacific Northwest National Laboratory, which is operated by Battelle for the U.S. Department of Energy under Contract No. DE-AC05-76RL01830.

%% file: basic/author_list_end.tex
\newcommand{\notds}{\nolinebreak\footnotemark\nolinebreak}
\renewcommand{\thefootnote}{$*$}

{
\onecolumn
\textbf{The DarkSide-20k Collaboration}
\footnotetext{Not a member of the DarkSide-20k Collaboration}

P.~Agnes\thanksref{Houston}\nolinebreak, 
S.~Albergo\thanksref{CTINFN}{CTUNI}\nolinebreak, 
I.~F.~M.~Albuquerque\thanksref{USP}\nolinebreak,
T.~Alexander\thanksref{PNNLaddress}\nolinebreak,
A.~Alici\thanksref{BOUniPHY}{BOINFN}\nolinebreak,
A.~K.~Alton\thanksref{Augustana}\nolinebreak,
P.~Amaudruz\thanksref{TRIUMFaddress}\nolinebreak,
M.~Arba\thanksref{CAINFN}\nolinebreak,
P.~Arpaia\thanksref{NAUniPHY}\nolinebreak,
S.~Arcelli\thanksref{BOUniPHY}{BOINFN}\nolinebreak
% ...

\begin{enumerate}[label=\textsuperscript{\arabic*}]
    \item {\label{Houston}\Houston}
    \item {\CTINFN \label{CTINFN}}
    \item {\CTUNI \label{CTUNI}}
    \item {\USP \label{USP}}
    \item {\PNNLaddress \label{PNNLaddress}}
    \item {\BOUniPHY \label{BOUniPHY}}
    \item {\BOINFN \label{BOINFN}}
    \item {\Augustana \label{Augustana}}
    \item {\TRIUMFaddress \label{TRIUMFaddress}}
    \item {\MendeleevUniverisity \label{MendeleevUniverisity}}
    \item {\LNLINFN \label{LNLINFN}}
    \item {\RHUL \label{RHUL}}
    \item {\MSU \label{MSU}}
    \item {\CIEMAT \label{CIEMAT}}
    \item {\CPPM \label{CPPM}}
    \item {\NAINFN \label{NAINFN}}
    \item {\PIINFN \label{PIINFN}}
    \item {\PIUniPHY \label{PIUniPHY}}
    \item {\BINP \label{BINP}}
    \item {\NSU \label{NSU}}
    \item {\CAINFN \label{CAINFN}}
    \item {\GEUni \label{GEUni}}
    \item {\GEINFN \label{GEINFN}}
    \item {\Carleton \label{Carleton}}
    \item {\CERNaddress \label{CERNaddress}}
    \item {\RMTreINFN \label{RMTreINFN}}
    \item {\RMTreUni \label{RMTreUni}}
    \item {\CAUniPHY \label{CAUniPHY}}
    \item {\AQLNGS \label{AQLNGS}}
    \item {\AQGSSI \label{AQGSSI}}
    \item {\CentroFermi \label{CentroFermi}}
    \item {\CAUniEEE \label{CAUniEEE}}
    \item {\NAUniPHY \label{NAUniPHY}}
    \item {\NAUniStruct \label{NAUniStruct}}
    \item {\OACINAF \label{OACINAF}}
    \item {\Zaragoza \label{Zaragoza}}
    \item {\Krakow \label{Krakow}}
    \item {\MIINFN \label{MIINFN}}
    \item {\TOINFN \label{TOINFN}}
    \item {\LPNHE \label{LPNHE}}
    \item {\RMUnoINFN \label{RMUnoINFN}}
    \item {\RMUnoUni \label{RMUnoUni}}
    \item {\SAUni \label{SAUni}}
    \item {\SAINFN \label{SAINFN}}
    \item {\MIPoliCHE \label{MIPoliCHE}}
    \item {\Petersburg \label{Petersburg}}
    \item {\Kurchatov \label{Kurchatov}}
    \item {\UMass \label{UMass}}
    \item {\UCDavis \label{UCDavis}}
    \item {\LNFINFN \label{LNFINFN}}
    \item {\APC \label{APC}}
    \item {\Princeton \label{Princeton}}
    \item {\ETHZ \label{ETHZ}}
    \item {\TOPoli \label{TOPoli}}
    \item {\WilliamsCollege \label{WilliamsCollege}}
    \item {\Hawaii \label{Hawaii}}
    \item {\TNFBK \label{TNFBK}}
    \item {\TNTIFPA \label{TNTIFPA}}
    \item {\UB \label{UB}}
    \item {\MEPhI \label{MEPhI}}
    \item {\JINR \label{JINR}}
    \item {\IHEPaddress \label{IHEPaddress}}
    \item {\ENUniCEE \label{ENUniCEE}}
    \item {\CTLNS \label{CTLNS}}
    \item {\Alberta \label{Alberta}}
    \item {\AstroCeNT \label{AstroCeNT}}
    \item {\FNALaddress \label{FNALaddress}}
    \item {\SNOLABaddress \label{SNOLABaddress}}
    \item {\Laurentian \label{Laurentian}}
    \item {\Temple \label{Temple}}
    \item {\Belgorod \label{Belgorod}}
    \item {\NAUniPHARM \label{NAUniPHARM}}
    \item {\ZaragozaARAID \label{ZaragozaARAID}}
    \item {\Queens \label{Queens}}
    \item {\Lancaster \label{Lancaster}}
    \item {\MIPoliICA \label{MIPoliICA}}
    \item {\Manchester \label{Manchester}}
    \item {\MIUni \label{MIUni}}
    \item {\BNLaddress \label{BNLaddress}}
    \item {\NAUniCHE \label{NAUniCHE}}
    \item {\VTech \label{VTech}}
    \item {\UCLA \label{UCLA}}
    \item {\FortLewis \label{FortLewis}}
    \item {\Lodz \label{Lodz}}
    \item {\CAUniCHE \label{CAUniCHE}}
    \item {\INCDTIM \label{INCDTIM}}
    \item {\ENEA \label{ENEA}}
    \item {\CS \label{CS}}
    \item {\Polaris \label{Polaris}}
    \item {\MISE \label{MISE}}
\end{enumerate}
}